\documentclass[11pt,a4paper]{article}

\usepackage[truedimen,margin=30mm]{geometry} 

\usepackage{mathrsfs}
\usepackage{amssymb}
\usepackage{amsmath}
\usepackage{ascmac}
\usepackage{amsthm}
\usepackage[dvipdfmx]{graphicx}
\usepackage{booktabs}
\usepackage{setspace}
\usepackage{natbib}
\usepackage[svgnames,dvipsnames,x11names]{xcolor}
\usepackage{bm}
\usepackage{url}
\usepackage{adjustbox}
\usepackage{times}

\usepackage{comment}
\usepackage{multicol,multirow}

\usepackage{titlesec}
\titleformat*{\section}{\large\bfseries}
\titleformat*{\subsection}{\it}

\newtheorem{thm}{Theorem}
\newtheorem{lem}{Lemma}

\newtheorem{algo}{Algorithm}

\def\ep{{\varepsilon}}

\def\beh{\widehat{\beta}}

\def\st{\widetilde{s}}

\def\mut{\widetilde{\mu}}
\def\Sit{\widetilde{\Sigma}}

\def\at{\widetilde{a}}
\def\bt{\widetilde{b}}
\def\pt{\widetilde{p}}
\def\mt{\widetilde{m}}
\def\St{\widetilde{S}}

\def\A{\mbox{\boldmath$A$}}
\def\B{\mbox{\boldmath$B$}}

\def\F{\mbox{\boldmath$F$}}
\def\G{\mbox{\boldmath$G$}}

\def\y{\mbox{\boldmath$y$}}
\def\f{\mbox{\boldmath$f$}}

\def\one{\mbox{\boldmath$1$}}

\def\bmu{\mbox{\boldmath$\mu$}}

\def\bSigma{\mbox{\boldmath$\Sigma$}}
\def\bbe{\mbox{\boldmath$\beta$}}
\def\bOm{\mbox{\boldmath$\Omega$}}
\def\bom{\mbox{\boldmath$\omega$}}

\def\barbeta{\overline{\beta}}

\title{{\bf Bayesian Spatial Predictive Synthesis  }}

\date{}
\author{}

\begin{document}

\maketitle
\doublespacing

\vspace{-1.6cm}
\begin{center}
{\large Danielle Cabel$^1$, Shonosuke Sugasawa$^2$, Masahiro Kato$^3$, \\
K\={o}saku Takanashi$^4$, and Kenichiro McAlinn$^1$}
\end{center}

\noindent
$^1$Fox School of Business, Temple University, Philadelphia, USA\\
$^2$Faculty of Economics, Keio University, Tokyo, Japan\\
$^3$Graduate School of Arts and Sciences, The University of Tokyo, Tokyo, Japan\\
$^4$Center for Advanced Intelligence Project, Riken, Tokyo, Japan\\

\vspace{-5mm}
\begin{center}
{\bf \large Abstract}
\end{center}

Due to spatial dependence-- often characterized as complex and non-linear-- model misspecification is a prevalent and critical issue in spatial data analysis and prediction.
As the data, and thus model performance, is heterogeneous, typical model selection and ensemble methods that assume homogeneity are not suitable.
We address the issue of model uncertainty for spatial data by proposing a novel Bayesian ensemble methodology that captures spatially-varying model uncertainty and performance heterogeneity of multiple spatial predictions, and synthesizes them for improved predictions, which we call Bayesian  spatial  predictive synthesis.
Our proposal is defined by specifying a latent factor spatially-varying coefficient model as the synthesis function, which enables spatial characteristics of each model to be learned and ensemble coefficients to vary over regions to achieve flexible predictions.
We derive our method from the theoretically best approximation of the data generating process, and show that it provides a finite sample theoretical guarantee for its predictive performance, specifically that the predictions are exact minimax.
Two MCMC strategies are implemented for full uncertainty quantification, as well as a variational inference strategy for fast point inference.
We also extend the estimation strategy for general responses.
Through simulation examples and two real data applications in real estate and ecology, our proposed Bayesian spatial predictive synthesis outperforms standard spatial models and ensemble methods, and advanced machine learning methods, in terms of predictive accuracy and uncertainty quantification, while maintaining interpretability of the prediction mechanism. 


\section{Introduction}
The modeling of spatial data-- data that are dispersed and linked to a geographical location-- has received considerable interest due to its abundance and relevance in numerous fields.
These data are characterized by their spatial dependence and correlation, where ``neighbors'' share features and may be clustered within certain regions, and taking into account these characteristics is critical to capture spatial heterogeneity and to predict unobserved locations \citep[see, e.g.,][for the different models used for this problem]{brunsdon1998geographically,anselin1988spatial,diggle1998model,gelfand2003spatial,wang2003generalized}.
Given the abundance and variety of spatial models, dealing with spatial model uncertainty is essential to achieve improved predictive accuracy and decision making.
One popular way to achieve this in practice is through ensembling multiple models. 

A notable drawback and limitation of existing approaches is the implicit assumption of spatial homogeneity, i.e., not taking into account the fact that data and performance are spatially heterogeneous, a defining feature of spatial data. 
For example, homogeneous model averaging methods have been considered for spatial autoregressive models \citep{debarsy2020bayesian,lesage2007bayesian,zhang2018spatial}, spatial error models \citep{greenaway2021spatial,liao2019spatial}, and hierarchical spatial models \citep{zhang2023exact}.
For a general ensemble of candidate models, stacking or super learner \citep{van2007super} is employed \citep{davies2016optimal}, but the optimal weights are still homogeneous, in the sense that it only provides one weight for each model for all locations. 

We contribute to this field by introducing a general framework to deal with model uncertainty in spatial data.
Our approach works within the framework of Bayesian predictive synthesis \citep[see, e.g.][]{mcalinn2019dynamic,mcalinn2020multivariate}, which is a coherent Bayesian framework for synthesizing multiple sources of information, based on agent opinion analysis \citep[see, e.g.][]{GenestSchervish1985,West1992c}.
Using this framework, we develop a Bayesian ensemble method for spatial data that explicitly takes into account the spatially dependent biases and dependencies among models, which we call Bayesian spatial predictive synthesis (BSPS).
Our proposed method departs from existing methods in a couple of critical ways.
First, predictions from spatial models are treated as latent factors, allowing for the synthesis model to learn model biases and dependencies. 
This is important, since, while model performances may be heterogeneous, the predictions will likely be dependent (have similar characteristics of heterogeneity), and they may be biased in similar or difference ways, depending on the region.
Since the models produce predictions individually (as if they are independent), it is important to learn these characteristics through data.
Second, the model weights are spatially varying to capture spatial heterogeneity of model importance.
Once the biases and dependencies are learnt, that information is then used to learn the model weights that reflect the spatial heterogeneity and model bias/dependence.
BSPS, thus, effectively learns the spatially varying coefficients that, in turn, improves predictive accuracy and decision making.

\begin{figure}[!tb]
\centering
\includegraphics[width=14cm,clip]{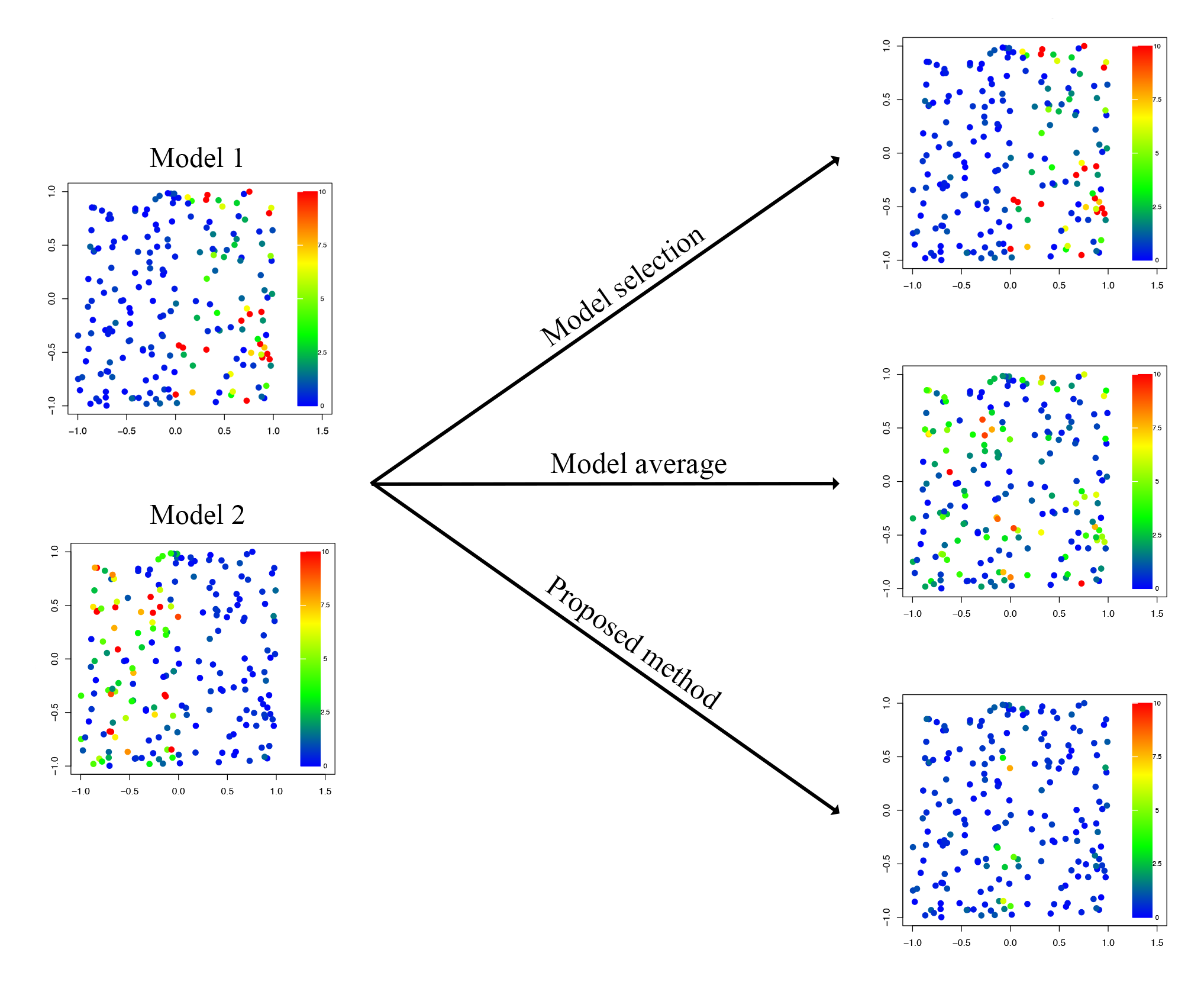}
\caption{Illustration of heterogeneous model performance for spatial data, and how model selection, simple model averaging, and our proposed BSPS performs. Each dot is the out of sample squared error.
\label{fig:spatial}
}
\end{figure}

To illustrate our motivation and contribution, consider a simple simulated data over a square (Figure~\ref{fig:spatial}).
In this illustration we have two models, Model 1 and 2, where the former performs well on the left, while the latter performs well on the right.
While this example is simple, it captures the essence of the spatial problem we are considering.
Specifically, the two regions can represent urban/suburban, mountainous/flat, or industrial/residential areas, where one model is good at modeling a certain type of area.
It is also important to note that the performance is not clearly separable, where there is mixed performance in the central region, as well as good predictions across the entire map.
This means that one cannot simply switch models for specific regions, since that ignores the good predictions in different regions, and how one divides a region is arbitrary and becomes increasingly difficult when there are many models that differ in their performance.
This problem is exacerbated when dealing with real world data, since these regions are often not well-defined.

In the right of the figure, we illustrate how different approaches produce different results.
If we were to select the best model, in this case Model 1, we effectively ignore the good performance of Model 2 in the right region.
Unless there is a model that uniformly outperforms all other models for all regions-- an assumption that cannot be made in most applications-- model selection will inevitably lose important information that can improve overall predictive performance.
If we were to average the two models (in this case with equal weights, a method often used in practice), we average out the performance of the two models, and the result is mixed.
This, in a different way, is losing critical information, as it ignores the performance heterogeneity and cannot leverage the fact that each model performs well in different regions.
Finally, our proposed approach learns the latent dependence and biases and leverages the spatial heterogeneity to produce improved performance over the entire map.
This simple illustration shows why spatially dependent ensemble methodologies that learns and leverages performance heterogeneity is critical in mitigating model uncertainty and improve performance.

While the motivation to develop spatially dependent ensemble methods is clear, the exact form of the method is not obvious.
In order to develop an ensemble method that is justifiable, with theoretical properties that are desirable, we first identify the best approximate model, under the assumption that the data generating process and predictions to be synthesized follow Gaussian processes.
The best approximate model, we show, is equivalent to a latent factor spatially varying coefficient model, used as a synthesis function, within the Bayesian predictive synthesis framework.
This is our proposed method, which we will expound in later sections.

Several computationally efficient algorithms are developed and implemented to produce posterior and predictive analysis, depending on the scale of the dataset.
The first two are MCMC based algorithms for full posterior analysis, one employing the nearest neighbor Gaussian process \citep{datta2016hierarchical}, which reduces the dimension for faster computation.
For larger datasets, we also develop a variational Bayes approximation \citep{blei2017variational} that produces even faster, accurate point predictions.
We also extend the estimation algorithm to deal with general responses, and specifically develop an efficient computation algorithm for the binary case, using the P\'{o}lya-gamma augmentation \citep{Polson2013}.

A series of simulated data and two real world applications, involving the occurrence of Tsuga canadensis and real-estate prices in Tokyo, Japan, illustrate the efficacy of our proposed method.
Through these applications, we show that our method has distinct advantages over competing methods, including statistical and machine learning methods for spatial data and ensemble methods.
Notably, we show that BSPS, synthesizing conventional spatial models, delivers better predictive accuracy than state-of-the-art machine learning methods, owing to the flexible ensemble BSPS provides through spatially varying model weights.

The rest of the article will proceed as follows. 
Section~\ref{sec:BPS} introduces a fundamental theory of BSPS by identifying the best approximate model used as a synthesis function, and shows that predictive distribution derived by BSPS is exact minimax. 
Details of the implementation of the proposed BSPS including its MCMC computational strategy, and extensions to general responses are given in Section~\ref{sec:implement}. 
We also develop alternative computational strategies for scalable inference.
Simulation studies are presented in Section~\ref{sec:sim}.  
Real world applications with the occurrence of Tsuga canadensis and apartment prices in Tokyo are presented in Section~\ref{sec:app}. 
The paper concludes with additional comments and closing remarks in Section~\ref{sec:conc}.
Further technical details and additional numerical results are presented in Supplementary Material.

\section{Bayesian Spatial Predictive Synthesis: Framework, Model, and Theory\label{sec:BPS}}

\subsection{General framework of Bayesian predictive synthesis}
Consider predicting a univariate outcome, $y(s)$, at some unobserved site, $s\in \mathcal{S}\subseteq \mathbb{R}^d$.
Suppose that a Bayesian decision maker, $\mathcal{D}$, uses information (predictive distributions) from $J$ models for $y(s)$, each of them denoted by the density function, $h_j(\cdot)$, for $j=1,\ldots,J$.
While $h_j(\cdot)$ can be any distribution, one pertinent example is a Gaussian predictive distribution, where the agent specifies the predictive (space-wise) mean and variance.
These forecast densities represent the individual inferences from the models and the collection of these forms the information set, $\mathcal{H}(s) = \{h_1(f_1(s)), \ldots, h_J(f_J(s))\}$, where $f_j(s)$ is a variable at site $s$.
Thus, formal Bayesian analysis indicates that $\mathcal{D}$ will predict $y(s)$ using its implied posterior predictive distribution, $p(y(s) | \mathcal{H}(s))$.
However, the set of $\mathcal{H}(s)$ is non-trivially complex, given its spatially varying structure of $J$ density functions. 
As these models are not ``independent''-- with information overlap among models-- there will be spatial dependencies and biases making straightforward Bayesian updating difficult.

The Bayesian predictive synthesis (BPS) framework \citep{mcalinn2019dynamic,GenestSchervish1985,West1992c,West1992d} provides a general and coherent way for Bayesian updating, given multiple predictive distributions.
Specifically, the Bayesian posterior is given as,
\begin{equation}\label{BPS}
\Pi_{\rm BPS}\big(y(s) | \Psi(s), \mathcal{H}(s)\big)=\int \alpha\big(y(s)|\f (s),\Psi(s)\big)\prod_{j=1}^J h_j\big(f_j(s)\big)df_j(s), 
\end{equation}
where $\alpha\big(y(s)|\f (s),\Psi(s)\big)$ is a synthesis function, $\Psi(s)$ represents the spatially varying parameters, and $\f (s)=(f_1(s),\ldots,f_J(s))$ is a vector of latent variables.  
Here, $\alpha\big(y(s)|\f (s),\Psi(s)\big)$ determines how the predictive distributions are synthesized and it includes a variety of existing combination methods, such as Bayesian model averaging and simple averaging \citep[e.g.][]{hoeting1999bayesian,geweke2011optimal,aastveit2018combined}, as special cases. 
The representation of (\ref{BPS}) does not require a full specification of the joint distribution of $y(s)$ and $\mathcal{H}(s)$, and it does not restrict the functional form of the synthesis function, $\alpha\big(y(s)|\f (s),\Psi(s)\big)$.
This allows $\mathcal{D}$ to flexibly specify how they want the information to be synthesized.
Note that (\ref{BPS}) is only a valid posterior if it satisfies the consistency condition. 
This condition states that, prior to observing $\mathcal{H}(s)$, $\mathcal{D}$ specifies their own prior predictive, $\Pi\{y(s)\}$, as well as their prior expectation of the model prediction, $\mathrm{E}[\prod_{j=1}^J h_j\big(f_j(s)\big)]$.
Then $\Pi\{y(s)\}=\int \alpha\big(y(s)|\f (s),\Psi(s)\big)\prod_{j=1}^J h_j\big(f_j(s)\big)df_j(s)$ must hold, meaning that the two priors that $\mathcal{D}$ specifies must be consistent with each other.

Treating the set of forecast densities as inherent latent factors linked to the outcome of interest, (\ref{BPS}) specifies a latent factor spatially varying  model, where the biases and dependencies of the agents are learned and updated as a function of location, to improve the overall, synthesized forecast \citep[see,][for further discussion]{mcalinn2019dynamic}.
Thus, even though the predictive densities are provided independently, as is the case in most applications, with optimism/pessimism and over/under-confidence, the BPS framework learns these features as latent states, given the data.

Since the BPS framework does not specify which model to use for the synthesis function, we derive a theoretical result that identifies a suitable model for spatial data to motivate our choice.
For this, we first formulate the data generating process (DGP) as a Gaussian field, which is flexible enough for many spatial applications.
Given this DGP, we identify a class of models-- spatially varying coefficient models-- that provides the best approximation, defined as the projection to the DGP that minimizes the MSE (Theorem~\ref{prp:best}).

\subsection{Specification of the best approximate synthesis function}
To cast the general representation (\ref{BPS}) to synthesize multiple spatial predictions, we need to identify a synthesis function, $\alpha\big(y(s)|\f(s),\Psi(s)\big)$,
that is justifiable. 
In this subsection, we derive the specific form of the synthesis function as the best approximation of the unknown data generating process.

We consider the task of predicting the data generating process, $y(s)$, with the predictive values from the $J$ models (predictive distributions), denoted by $\f(s)=(f_{1}(s),\ldots,f_{J}(s))$.
Following the framework of BPS (\ref{BPS}), we assume that $f_1,\ldots,f_J$ are mutually independent, and define $a_j(s)=\mathbb{E}[f_j(s)]$ and $b_j(s)={\rm Var}(f_j(s))$. 
For fixed $s$, we consider predicting $y(s)$ by the form $\mu(s)=\beta_0+\sum_{j=1}^J \beta_j f_j(s)$, a linear combination of $f_j(s)$.
The optimal coefficients, $\beta_0^{\ast},\beta_1^{\ast},\ldots,\beta_J^{\ast})$ that minimize the expected squared error, $\mathbb{E}[\{y(s)-\mu(s)\}^2]$ can be expressed as $\beta_j^{\ast}=\mathbb{E}[\{f_j(s)-a_j(s)\}y(s)]/b_j(s)$ for $j=1,\ldots,J$ and $\beta_0^{\ast}=\mathbb{E}[y(s)]-\sum_{j=1}^J \beta_j^{\ast}a_j(s)$.
Then, under a general situation where the expectation $a_j(s)$ and variance $b_j(s)$ exhibits spatial nonstationarity, the optimal coefficient $\beta_j^{\ast}$ would also depend on $s$.
This means that the coefficients (model weights) should be spatially varying to best approximate the unknown DGP using multiple models. 
We summarize the statement in the following theorem: 

\begin{thm}\label{prp:best} 
Given the prediction models as random variables, $(f_{1}(s),\ldots,f_{J}(s))$, the best linear approximation model to the DGP $y(s)$ can be expressed as 
\begin{equation}\label{eq:SVCM_Apprx}
y(s)=\beta_{0}(s)+\sum_{j=1}^{J}\beta_{j}(s)f_{j}(s)+\varepsilon(s),
\end{equation}
where $\beta_{0}(s),\beta_{1}(s),\ldots,\beta_{J}(s)$ are unknown spatially varying coefficients, and $\varepsilon(s)$ is an error term satisfying $\mathbb{E}[\varepsilon(s)]=0$. 
\end{thm}

From Theorem~\ref{prp:best}, it follows that using (\ref{eq:SVCM_Apprx}) as the synthesis function in (\ref{BPS}) obtains the best approximation of the data generating process given multiple prediction models.
Throughout this paper, we denote BPS that uses the synthesis function (\ref{eq:SVCM_Apprx}) in (\ref{BPS}),
\textit{Bayesian spatial predictive synthesis (BSPS)}.
In practice, $\beta_j(s)$ is unknown and will be estimated from the observed data. 
In particular, according to the above discussion, the coefficient $\beta_j(s)$ defined as the moments of $f_j(s)$ and $y(s)$, $\beta_(s)$ would be smoothly varying over the space.
This motivates the use of Gaussian processes for estimating $\beta_(s)$, which will be discussed in the next section.

Regarding the approximate model (\ref{prp:best}), we also provide some insights into the theoretical properties of BSBS.
In particular, we can show that the predictive distribution induced from (\ref{prp:best}) is exact minimax (i.e., minimax in finite sample) under some conditions.
The details are provided in Supplementary Material~\ref{sec:minimax}.

\section{Implementation of BSPS}
\label{sec:implement}

\subsection{Synthesis model, posterior computation and spatial prediction}
\label{sec:MCMC}

To synthesize multiple spatial predictions, we fit the synthesis model (\ref{eq:SVCM_Apprx}) to the observed data. 
We employ a Gaussian process to estimate unknown synthesis coefficients $\beta_j(s)$, which are independent for $j=0,\ldots,J$, namely $\beta_j(s)\sim {\rm GP}(\barbeta_j, \theta_j)$, where ${\rm GP}(\barbeta_j, \theta_j)$ denotes a Gaussian process with mean $\barbeta_j$ and covariance parameters $\theta_j$. 
In what follows, we assume that $\barbeta_j$ is fixed and $\theta_j=(\tau_j,g_j)$, where $\tau_j$ and $g_j$ are unknown scale and spatial range parameters. 
A reasonable choice is $\barbeta_j=1/J$, meaning that the prior synthesis is simple averaging of $J$ prediction models for all the locations, so we use $\barbeta_j=1/J$ as a default choice.

Suppose we observe samples at $n$ locations, $s_1,\ldots,s_n \in \mathcal{S}$.
Let $y_i=y(s_i)$, $f_{ji}=f_j(s_i)$, $\ep_i=\ep(s_i)$, and $\beta_{ji}=\beta_j(s_i)$.
Then, the model in (\ref{eq:SVCM_Apprx}) at the sampled locations is written as 
\begin{equation}\label{SVC2}
\begin{split}
&y_i=\beta_{0i} + \sum_{j=1}^J\beta_{ji}f_{ji} + \ep_i, \ \ \  \ep_i\sim N(0,\sigma^2), \ \ \ i=1,\ldots,n,\\
&\bbe_j\equiv (\beta_{j1},\ldots,\beta_{jn})^\top\sim N(\barbeta_j \one_n,\tau_j \G(g_j)), \ \  j=0,\ldots,J,
\end{split}
\end{equation}
where the $(i,i')$-element of $\G(g_j)$ is $C(\|s_i-s_{i'}\|; g_j)$ with valid correlation function $C(\cdot; g_j)$ and spatial range parameter $g_j$ as defined above. 
The model in (\ref{eq:SVCM_Apprx}) is quite similar to the spatially varying coefficient model \citep{gelfand2003spatial}, but the difference is that the latent factor $f_j(s)$ in (\ref{eq:SVCM_Apprx}) is a random variable rather than fixed covariates, as in the standard varying coefficient model.
For the prior distributions of the unknown parameters, we use $\sigma^2\sim {\rm IG}(a_\sigma,b_\sigma)$, $\tau_j\sim {\rm IG}(a_\tau,b_\tau)$, and $g_j\sim U(\underline{g}, \overline{g})$, independently for $j=1,\ldots,J$.
We obtain the joint posterior distribution
\begin{align*}
&\pi(\sigma^2)\prod_{j=0}^J\pi(\tau_j)\pi(g_j)\phi_n(\bbe_j; \barbeta_j \one_n, \tau_j \G(g_j)) \times \prod_{j=1}^J\prod_{i=1}^n g_j(f_{ij}) 
\times \prod_{i=1}^n \phi\Big(y_i; \beta_{0i} + \sum_{j=1}^J\beta_{ji}f_{ji}, \sigma^2\Big)
\end{align*}
where $\pi(\sigma^2), \pi(\tau_j)$ and $\pi(g_j)$ are prior distributions, and $\phi_n(\cdot; \bmu, \bSigma)$ denotes a $n$-dimensional normal distribution with mean vector $\bmu$ and covariance matrix $\bSigma$.

At location, $s$, the BSPS analysis will include inferences on the latent factor states, $f_j(s)$, as well as the spatially varying BSPS model parameters $\Psi(s)$. 
We first provide a computation algorithm using Markov chain Monte Carlo (MCMC). 
Suppose that $f_{ji}\sim N(a_{ji}, b_{ji})$ is received independently for $j=1,\ldots,J$ and $i=1,\ldots,n$, where $a_{ji}$ and $b_{ji}$ are provided by the $J$ models. 
The MCMC algorithm to generate posterior samples of $\{f_{ji}\}, \{\beta_j\}, \{\tau_j\}, \{g_j\}$ and $\sigma^2$ is given as follows:

\begin{itemize}
\item[-]
(Sampling of $f_{ji}$) \ \ \ 
Generate $f_{ji}$ from $N(A_{ji}^{(f)}B_{ji}^{(f)}, A_{ji}^{(f)})$, where 
$$
A_{ji}^{(f)}=\left(\frac{\beta_{ji}^2}{\sigma^2}+\frac1{b_{ji}}\right)^{-1}, \ \ \ \ \ 
B_{ji}^{(f)}=\frac{\beta_{ji}}{\sigma^2}\bigg(y_i-\beta_{0i}-\sum_{k\neq j}\beta_{ki}f_{ki}\bigg)+\frac{a_{ji}}{b_{ji}}
$$

\item[-]
(Sampling of $\bbe_j$) \ \ 
Generate $\bbe_j$ from $N(\A_j^{(\beta)}\B_j^{(\beta)}, \A_j^{(\beta)})$, where 
\begin{align*}
&\A_j^{(\beta)}=\big\{\sigma^{-2}\bOm_j + \tau_j^{-1}\G(g_j)^{-1}\big\}^{-1}, \\
&\B_j^{(\beta)}=\frac1{\sigma^2}\f_j\circ\bigg(y-\bbe_0-\sum_{k\neq j}\f_k\circ \bbe_k\bigg)+\frac{\barbeta_j}{\tau_j}\G(g_j)^{-1}\one_n,
\end{align*}
with $\bOm_j={\rm diag}(f_{j1}^2,\ldots,f_{jn}^2)$ and $\f_j=(f_{j1},\ldots,f_{jn})$.
Note that $\circ$ denotes the Hadamard product.

\item[-] 
(Sampling of $\tau_j$) \ \ Generate $\tau_j$ from ${\rm IG}(a_\tau+n/2,b_{\tau}+(\bbe_j-\barbeta_j\one_n)^\top \G(g_j)^{-1}(\bbe_j-\barbeta_j\one_n)/2)$.

\item[-] 
(Sampling of $g_j$) \ \ 
The full conditional of $g_j$ is proportional to 
$$
|\G(g_j)|^{-1/2}\exp\Big\{-\frac1{2\tau_j}(\bbe_j-\barbeta_j\one_n)^\top \G(g_j)^{-1}(\bbe_j-\barbeta_j\one_n)\Big\}, \ \ \ \ g_j\in (\underline{g}, \overline{g}).
$$
A random-walk Metropolis-Hastings is used to sample from this distribution. 

\item[-] 
(Sampling of $\sigma^2$) \ \ Using the conditionally conjugate prior $\sigma^2\sim {\rm IG}(a_\sigma,b_{\sigma})$, the full conditional is $\sigma^2\sim {\rm IG}(a_\sigma+n/2,b_{\sigma}+\sum_{i=1}^n(y_i-\beta_{0i}-\sum_{j=1}^J\beta_{ji}f_{ji})^2/2)$.
\end{itemize}

Each item is sampled for $j=1,\ldots,J$, per MCMC iteration. 
The information is then updated with each iteration to be used throughout the algorithm. 
Note that, in practice, $h_j(\cdot)$ is very likely to be a conditional density depending on some covariates. 
Extension to such a case is trivial.

Turning to predictions, let $s_{n+1}$ be a new location where we are interested in predicting $y_{n+1}\equiv y(s_{n+1})$, assuming that the predictive distributions of $\f_{n+1}=(f_1(s_{n+1}),\ldots,f_J(s_{n+1}))$, namely, predictive distributions of the $J$ models, are available.
Then, the predictive distribution of $y_{n+1}$ is obtained as 
\begin{equation}\label{pred-dist}
\begin{split}
p(y_{n+1}|\y, \f_{n+1}) 
&=\int \phi\Big(y_{n+1}; \beta_{0,n+1} + \sum_{j=1}^J\beta_{j,n+1}f_{j,n+1}), \sigma^2\Big)
\prod_{j=1}^J h_j(f_{j,n+1}) df_{j,n+1}
\\
& \ \ \ \ \ \ 
\times \prod_{j=0}^J p(\beta_{j,n+1}|\bbe_j; \tau_j, g_j)d\beta_{j,n+1} \times \pi(\Theta|\y)d\Theta,
\end{split}
\end{equation}
where $\Theta$ is a collection of $\{f_{ji}\}, \{\beta_j\}, \{\tau_j\}, \{g_j\}$ and $\sigma^2$, $p(\beta_{j,n+1}|\bbe_j; \tau_j, g_j)$ is the conditional distribution of $\beta_{j,n+1}$ given $\bbe_j$, and $\pi(\Theta|\y)$ is the posterior distribution of $\Theta$.
Under the assumption of Gaussian process on $\beta_j(s)$, the conditional distribution of $\beta_{j,n+1}$ is given by 
$$
N(\G_{n+1}(g_j)^{\top}\G(g_j)^{-1}\bbe_j, \{\tau_j- \tau_j\G_{n+1}(g_j)^{\top}\G(g_j)^{-1}\G_{n+1}(g_j)\}^{-1}),
$$
where $\G_{n+1}(g_j)=(C(\|s_{n+1}-s_1\|;g_j),\ldots,C(\|s_{n+1}-s_n\|;g_j))^\top$.
Sampling from the predictive distribution (\ref{pred-dist}) can be easily carried out by using the posterior samples of $\Theta$. 
First, independently generate $f_j(s_{n+1})$ from the predictive distribution of the $j$th model and generate $\beta_{j,n+1}$ from its conditional distribution given $\Theta$.
Then, we can generate $y_{n+1}$ from $N(\beta_{0,n+1} + \sum_{j=1}^J\beta_{j,n+1}f_{j,n+1}, \sigma^2)$.

The full Gaussian process is known to be computationally prohibitive under large spatial data, since it requires computational cost $O(Jn^3)$ for each MCMC iteration of BSPS. 
To overcome the difficulty, we employ an $m$-nearest neighbor Gaussian process \citep{datta2016hierarchical} for $\beta_j(s)$, which uses a multivariate normal distribution with a sparse precision matrix for $\beta_j(s_1),\ldots,\beta_j(s_n)$.
Then, the computational cost at each iteration is reduced to $O(Jnm^2)$, which is a drastic reduction from the original computation cost $O(Jn^3)$, since $m$ can be set to a small value (e.g. $m=10$), even under $n\approx 10^4$.
The detailed sampling steps under the nearest neighbor Gaussian process are provided in Supplementary Material~\ref{sec:NNGP-pos}.

\subsection{Variational Bayes approximation for fast point prediction synthesis}\label{sec:VB}
While the MCMC algorithm does provide full posterior estimation, it can also be prohibitively slow when the number of sampled locations or predictors is large.
As such, we also develop an approximation algorithm using mean field variational Bayes (MFVB) approximation that is significantly more efficient than its MCMC counterpart. 
In applying the MFVB approximation, we assume that the prior distributions of the spatial range parameters, $g_0,g_1,\ldots,g_J$, are the uniform distribution on $\{\eta_1,\ldots,\eta_L\}$.
The MFVB approximates the posterior distributions through the form
$$
q(\{f_{ji}\}, \{\bbe_j\}, \{\tau_j\}, \{g_j\},\sigma^2)=q(\sigma^2)\prod_{j=0}^J q(\bbe_j)q(\tau_j)q(g_j)\prod_{i=1}^n q(f_{ji}),
$$
and each variational posterior can be iteratively updated by computing, for example, $q(\bbe_j)\propto \exp(E_{-\beta_j}[\log p(y,\Theta)])$, where $\Theta=(\{f_{ji}\}, \{\bbe_j\}, \{\tau_j\}, \{g_j\},\sigma^2)$, and $E_{-\beta_j}$ denotes the expectation with respect to the marginal variational posterior of the parameters other than $\bbe_j$.  
From the forms of full conditional posterior distributions given in Section \ref{sec:MCMC}, the following distributions can be used as variational distributions: 
\begin{align*}
&q(f_{ji}) \sim N(\mt_{ji}, \st^2_{ji}), \ \ \ 
q(\bbe_j)\sim N(\mut_{j}, \Sit_{j}), \ \ \ 
q(\tau_j)\sim {\rm IG}(\at_{\tau_j}, \bt_{\tau_j}), \\
&q(g_j)\sim {\rm D}(\pt_{j1},\ldots,\pt_{jL}), \ \ \ \ 
q(\sigma^2)\sim {\rm IG}(\at_{\sigma}, \bt_{\sigma}), 
\end{align*}
where ${\rm D}(\pt_{j1},\ldots,\pt_{jL})$ is a discrete distribution on $\{\eta_1,\ldots,\eta_L\}$, such that $P(g_j=\eta_\ell)=\pt_{j\ell}$.
The MFVB algorithm is described as follows:

\begin{algo}\label{algo:VB}
{\small 
Starting with $\mt_{ji}^{(0)}$, $\st_{ji}^{2(0)}$, $\mut_{j}^{(0)}, \Sit_{j}^{(0)}$, $\at_{\tau_j}^{(0)}, \bt_{\tau_j}^{(0)}$, $\pt_{j\ell}^{(0)}$, $\at_{\sigma}^{(0)}, \bt_{\sigma}^{(0)}$ and $r=0$, repeat the following process until numerical convergence:
for $j=1,\ldots,J$, update $\mt_{ji}$ and $\st_{ji}^{2}$ as 
\begin{align*}
&\st_{ji}^{2(t+1)}\leftarrow 
\left\{\frac1{b_{ji}}+(\mut_{ji}^{(t)2}+\Sit_{jii}^{(t)})\frac{\at_{\sigma}^{(t)}}{\bt_{\sigma}^{(t)}}\right\}^{-1}, \\
&\mt_{ji}^{(t+1)}\leftarrow 
\frac{a_{ji}}{b_{ji}}+\mut_{ji}^{(t)}\frac{\at_{\sigma}^{(t)}}{\bt_{\sigma}^{(t)}}\bigg(y_i-\mut_{0i}^{(t)}
-\sum_{k<j}\mut_{ki}^{(t)}\mt_{ki}^{(t+1)}
-\sum_{k>j}\mut_{ki}^{(t)}\mt_{ki}^{(t)}\bigg)
\Big/\st_{ji}^{2(t+1)}.
\end{align*}
For $j=1,\ldots,J$, update $\mut_j$ and $\Sit_{j}$ as 
\begin{align*}
&\Sit_j^{(t+1)}\leftarrow
\left\{\Omega_j^{(t+1)}\frac{\at_{\sigma}^{(t)}}{\bt_{\sigma}^{(t)}} + \sum_{\ell=1}^L \pt_{j\ell}^{(t)}G(\eta_\ell)^{-1}\frac{\at_{\tau_j}^{(t)}}{\bt_{\tau_j}^{(t)}}\right\}^{-1}, \\
&\mut_j^{(t+1)}\leftarrow
\big\{\Sit_j^{(t+1)}\big\}^{-1}\frac{\at_{\sigma}^{(t)}}{\bt_{\sigma}^{(t)}} \mt_j^{(t+1)}\circ
\bigg(y-\mut_0^{(t+1)}
-\sum_{k<j}\mut_k^{(t+1)}\circ \mt_k^{(t+1)}
-\sum_{k>j}\mut_k^{(t)}\circ \mt_k^{(t+1)}
\bigg),
\end{align*}
where $\Omega_j^{(t+1)}={\rm diag}(\mt_{j1}^{(t+1)2}+\st_{j1}^{2(t+1)},\ldots,\mt_{jn}^{(t+1)2}+\st_{jn}^{2(t+1)})$.
For $j=0,\ldots,J$, set $\at_{\tau_j}^{(t+1)}=a_{\tau}+n/2$ and update $\bt_{\tau_j}$ as 
\begin{align*}
\bt_{\tau_j}^{(t+1)} \leftarrow 
b_\tau + \frac12{\rm tr}\left\{\left(\mut_j^{(t+1)}\mut_j^{(t+1)\top} + \Sit_j^{(t+1)}\right)\sum_{\ell=1}^L\pt_{j\ell}^{(t)}G(\eta_\ell)^{-1}\right\}.
\end{align*}
For $j=0,\ldots,J$, update $\pt_{j\ell}$ as  
$$
\pt_{j\ell}^{(t+1)} \leftarrow 
\frac{|G(\eta_\ell)|^{-1/2}\exp\left(-\at_{\tau_j}^{(t+1)} {\rm tr}\left\{(\mut_j^{(t+1)} \mut_j^{(t+1)\top}  + \Sit_j^{(t+1)} )G(\eta_\ell)^{-1}\right\}/2\bt_{\tau_j}^{(t+1)} \right)}
{\sum_{\ell'=1}^L|G(\eta_{\ell'})|^{-1/2}\exp\left(-\at_{\tau_j}^{(t+1)} {\rm tr}\left\{(\mut_j^{(t+1)} \mut_j^{(t+1)\top}  + \Sit_j^{(t+1)} )G(\eta_{\ell'})^{-1}\right\}/2\bt_{\tau_j}^{(t+1)} \right)}.
$$
Set $\at_{\sigma}^{(t+1)}=a_{\sigma}+n/2$ and update $\bt_{\sigma}$ as
\begin{align*}
\bt_{\sigma}^{(t+1)} \leftarrow &
\bigg(y-\mut_0^{(t+1)}-\sum_{j=1}^J\mut_j^{(t+1)}\circ \mt_j^{(t+1)}\bigg)^\top \bigg(y-\mut_0^{(t+1)}-\sum_{j=1}^J\mut_j^{(t+1)}\circ \mt_j^{(t+1)}\bigg)
+
{\rm tr}(\Sit_0^{(t+1)}) \\
& + \sum_{j=1}^J {\rm tr}\left\{(\mut_j^{(t+1)}\mut_j^{(t+1)\top}  + \Sit_j^{(t+1)})\circ (\mt_j^{(t+1)}\mt_j^{(t+1)\top}+\St_j^{(t+1)})\right\}  - \sum_{i=1}^n\sum_{j=1}^J \mt_{ji}^{(t+1)2}\mut_{ji}^{(t+1)2}.
\end{align*}
}
\end{algo}

\medskip
A reasonable starting value for Algorithm \ref{algo:VB} is the posterior mean of a small number of MCMC samples.
We note that the updating step may contain calculations of the inverse of $n\times n$ matrices, as in the MCMC algorithm, which could be computationally prohibitive when $n$ is large. 
Alternatively, we can also develop a variational approximation algorithm for the nearest neighbor Gaussian process.

\subsection{BSPS under general types of response variables}
The proposed BSPS framework (\ref{BPS}) can be extended to situations with general types of outcomes, by generalizing the synthesis model (\ref{eq:SVCM_Apprx}) to generalized spatially varying models \citep[e.g.][]{gelfand2003spatial,kim2021generalized}. 
Here, we consider a specific situation where $y_i$ is a binary response, which will be treated in Section~\ref{sec:app-binary}.
The linear latent factor model (\ref{SVC2}) for continuous response can be modified as 
\begin{equation}\label{BPS-bin}
y_i|\psi_i\sim {\rm Ber}\left(\frac{\exp(\psi_i )}{1+\exp(\psi_i)}\right),  \ \ \ \ 
\psi_i=\beta_{0i}+\sum_{j=1}^J \beta_{ji}f_{ji}, \ \ \ i=1,\ldots,n
\end{equation}
where $\beta_{ji}$ follows the same Gaussian process given in (\ref{SVC2}).
Suppose that $f_{ji}$ is a predictive (posterior) distribution of logit-transformed probability, and assume that $f_{ji}\sim N(a_{ji}, b_{ji})$ with fixed $a_{ji}$ and $b_{ji}$.

To enhance the efficiency of posterior computation, we employ the following P\'{o}lya-gamma data augmentation \citep{Polson2013}:
$$
\frac{\exp(\psi_i y_i)}{1+\exp(\psi_i)}
=\frac12\exp\left\{\left(y_i-\frac12\right)\psi_i\right\}\int_0^{\infty}\exp\left(-\frac12\omega_i\psi_i^2\right)p(\omega_i; 1, 0)d\omega_i,
$$
where $p(\cdot; b, c)$ denotes the P\'{o}lya-gamma density with parameters $b$ and $c$.
Then, the full conditional distribution of $\bbe_j\ (j=0,\ldots,J)$ is a normal distribution, $N(\A_j^{(\beta)}\B_j^{(\beta)}, \A_j^{(\beta)})$, where
\begin{align*}
\A_j^{(\beta)}&=\Big\{{\rm diag}(\omega_1f_{j1}^2,\ldots,\omega_n f_{jn}^2) + \tau_j^{-1}\G(g_j)^{-1}\Big\}^{-1}, \\
\B_j^{(\beta)}&=\f_j\circ \Big\{\y^{\ast}-\bom\circ \Big(\bbe_0+\sum_{k\neq j}\f_k\circ \bbe_k\Big)\Big\}+\frac{\barbeta_j}{\tau_j}\G(g_j)^{-1}\one_n,
\end{align*}
where $\y^{\ast}=(y_1-1/2,\ldots,y_n-1/2)$ and $\bom=(\omega_1,\ldots,\omega_n)$.
The full conditional distribution of $f_{ji}$ is $N(A_{ji}^{(f)}B_{ji}^{(f)}, A_{ji}^{(f)})$, where 
$$
A_{ji}^{(f)}=\bigg(\omega_i\beta_{ji}^2 + \frac{1}{a_{ji}}\bigg)^{-1}, \ \ \ \ 
B_{ji}^{(f)}=\beta_{ji}\bigg\{\Big(y_i-\frac12\Big)-\omega_i\bigg(\beta_{0i}+\sum_{k\neq j}\beta_{ki}f_{ki}\bigg)\bigg\}+\frac{b_{ji}}{a_{ji}}.
$$
Finally, the full conditional distribution of $\omega_i$ is ${\rm PG}(1, \psi_i)$.

\section{Simulation Studies \label{sec:sim}}
This section provides a simulation study to illustrate the efficacy of our proposed BSPS compared to other methods for spatial data.

\subsection{Empirical behavior of BSPS}
\label{sec:emp_behavior}
We first illustrate how our proposed BSPS synthesizes candidate models. 
We set $n=300$ (training sample size) and generated two-dimensional location information $s_i$ (for $i=1,\ldots,n$) from the uniform distribution on $[-1, 1]^2$.
Let $z_1(s_i)$ and $z_2(s_i)$ be the two independent realizations of a spatial Gaussian process with mean zero and a covariance matrix defined from an isotropic exponential function: ${\rm Cov}(z_k(s_i), z_k(s_j))=\exp(-\|s_i-s_j\|/0.5)$ for $k=1,2$.  
Then, we define two covariates $x_1(s_i)$ and $x_2(s_i)$ via linear transformations, $x_1(s_i)=z_1(s_i)$ and $x_2(s_i)=rz_1(s_i)+\sqrt{1-r^2}z_2(s_i)$ with $r=0.2$, which allows dependence between $x_1(s_i)$ and $x_2(s_i)$. 
The response variable $y(s_i)$ at each location is generated from the following process: 
$$
y(s_i) = 
\begin{cases}
w(s_i)+x_1(s_i) -0.5x_2^2(s_i) + \ep(s_i), & s_i\in D_1,\\
w(s_i)+x_1^2(s_i) + x_2^2(s_i) + \ep(s_i), & s_i\in D_2.
\end{cases}
$$
where $D_1=\{s_i=(s_{i1},s_{i2})\ |\  s_{i1}\leq 0\}$ and $D_2=\{s_i=(s_{i1},s_{i2})\ | \ s_{i1}> 0\}$. 
Here $w(s_i)$ is a spatial random effect following a mean-zero Gaussian process with spatial covariance function, ${\rm Cov}(s_i, s_{i'})=(0.3)^2\exp(-\|s_i-s_{i'}\|/0.3)$, and $\ep(s_i)$ is an independent error term distributed as $\ep(s_i)\sim N(0,1)$.
Note that, in the above setting, the spatial region is divided into two sub-regions, where the mean structure of the response, as a function of covariates, is different.
For the data generated from the process, we apply a quadratic regression model without spatial effects, $y(s_i)=\beta_0+\beta_1x_1(s_i)+\beta_2x_1(s_i)^2+\beta_3x_2(s_i)+\beta_4x_2(s_i)^2+\ep_i,$ to subsamples in $D_1$ (denoted by QR1) and compute the predictive mean and variance of all the samples, which are denoted by $a_{1i}$ (mean) and $b_{1i}$ (variance), respectively. 
We conduct the same procedure using subsamples in $D_2$ (denoted by QR2) to obtain $a_{2i}$ and $b_{2i}$.
We apply BSPS using the two prediction models, $f_{ji}\sim N(a_{ji},b_{ji})$ with $j=1,2$, and an exponential kernel, $(G(g_j))_{ii'}=\exp(\|s_i-s_{i'}\|/g_j)$ in 10-nearest neighbor Gaussian processes. 
We generate 1000 posterior samples after discarding the first 1000 samples as burn-in.

We first evaluate the predictive performance in non-sampled locations. 
We generated 200 additional locations, as with $w(s_i)$, $x_1(s_i)$, $x_2(s_i)$, and $y(s_i)$, according to the same data generating process.
Using the generated posterior samples from BSPS, posterior samples of the spatially varying coefficients in the non-sampled locations were generated to get posterior predictive distributions of the response in non-sampled locations. 
We evaluate the mean squared error (MSE) of the posterior predictive means of BSPS, as well as predictors of the two quadratic regressions. 
For comparison, we employ two methods of prediction synthesis, Bayesian model averaging (BMA) and simple averaging (SA). 
In the former method, we compute the Bayesian information criterion for the two quadratic models to approximate the marginal likelihood.
In the latter method, the prediction results of QR1 and QR2 are simply averaged. 
The MSE of non-sampled locations are $1.33$ (BSPS), $2.61$ (BMA), and $4.19$ (SA), while the MSE of the two models are $2.61$ (QR1) and $10.85$ (QR2). 
Since QR1 is estimated using data only in $D_1$, its predictive performance in $D_2$ is not expected to be good, due to the difference in true regression structures between $D_1$ and $D_2$, which leads to QR1 having a large MSE.
The same explanations can be given for QR2 and its MSE. 
We found that the model weight for QR1 in BMA is almost $1$, so  the performance of BMA and QR1 is almost identical. 
It is reasonable that SA performs worse than QR1, since it gives equal weight to QR2, which does not perform well in this example. 
Comparatively, BSPS provides much better prediction results than the two ensemble methods in terms of MSE. 
The main reason is that BSPS can combine the two models with spatially varying weights, and such design of weights is essential in this example, since the usefulness of the two models are drastically different in $D_1$ and $D_2$. 
Furthermore, although both QR1 and QR2 do not take into account the existence of the spatial random effect, the intercept term in BSPS could successfully capture the remaining spatial variation, which increases the prediction accuracy in this example.

\begin{figure}[!tb]
\centering
\includegraphics[width=14cm,clip]{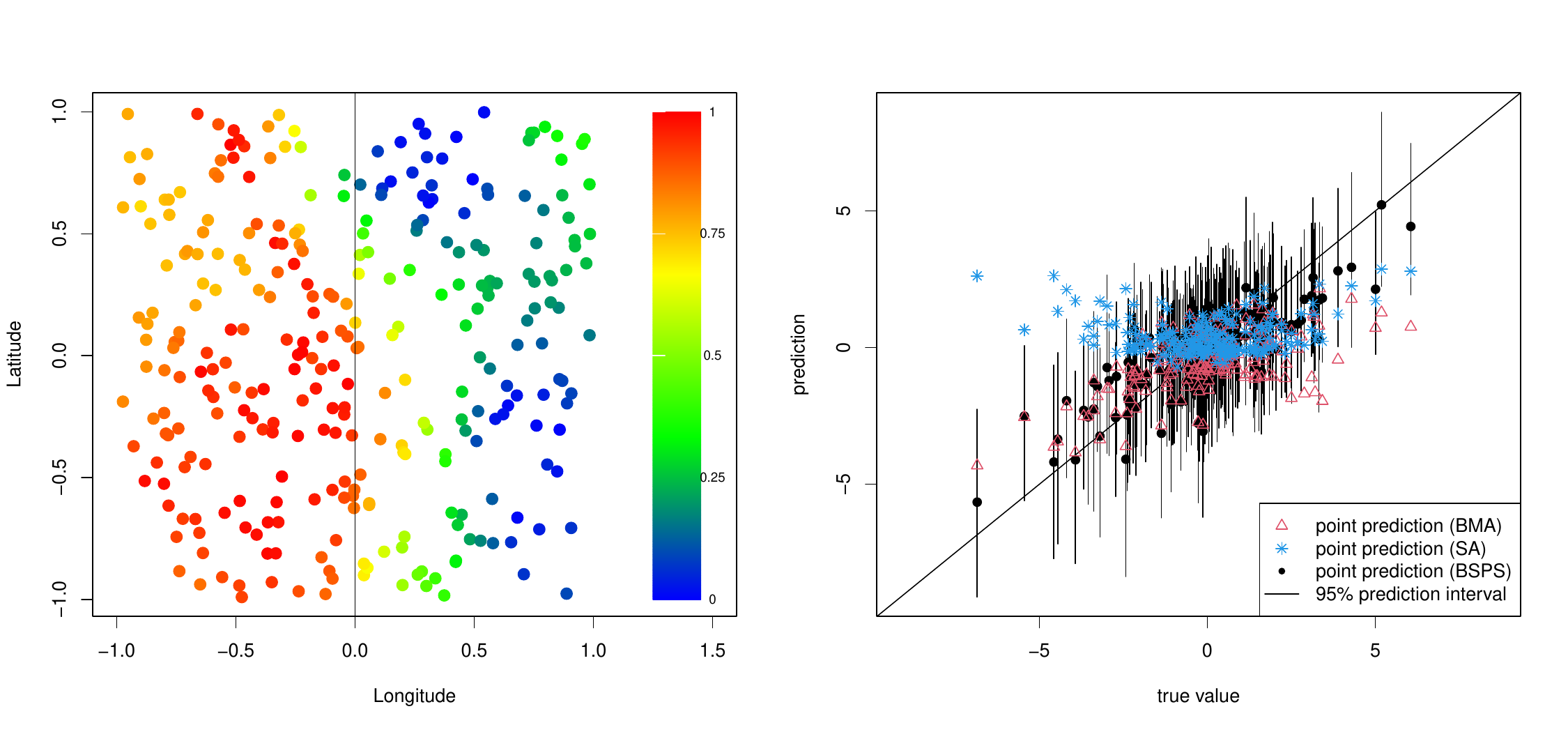}
\caption{Left: Spatial plot of the ratio of two coefficients, $|\beh_{1i}|/(|\beh_{1i}|+|\beh_{2i}|)$.  
Right: point prediction made by BMA, SA, and BSPS, and $95\%$ prediction intervals (vertical lines) obtained from the predictive distributions of BSPS. 
\label{fig:sim-toy}
}
\end{figure}

To see how BSPS works in this example, we compute the ratio of two coefficients, $|\beh_{1i}|/(|\beh_{1i}|+|\beh_{2i}|)$, where $\beh_{1i}$ and $\beh_{2i}$ are posterior means of $\beta_{1i}$ (weight for QR1) and $\beta_{2i}$ (weight for QR2), which shows the importance of the prediction made by QR1. 
The result is shown in the left panel of Figure \ref{fig:sim-toy}, which clearly shows that the model weight for QR1 is large in $D_1$ (left region) and is close to 0 in $D_2$, where QR1 is not expected to predict well. 
This means that BSPS can automatically detect the effective model at local regions through Bayesian updating.
We also note that the model weight smoothly changes over the region and two prediction models.

We evaluate the coverage accuracy of the $95\%$ interval prediction. 
In the right panel of Figure \ref{fig:sim-toy}, we present the mean values (point prediction) of predictive distributions with their associated $95\%$ prediction intervals.
The result shows that the prediction intervals mostly cover the true values with reasonable interval lengths. 
The coverage proportion is 0.97, which is practically equivalent to the nominal level, illustrating how well-calibrated BSPS is.
For comparison, we also present the point prediction made by BMA and SA in the right panel of Figure \ref{fig:sim-toy}.
It can be seen that BMA and SA fail to predict observations having large absolute values.

We next consider different prediction models. 
In addition to QR1 and QR2, we employ spatial regression (SPR) with quadratic terms and spatial effects modeled by the 10-nearest neighbor Gaussian process and an additive model (AM) without spatial effects. 
We then consider synthesizing QR1, QR2, and SPR (denoted by BSPS-ad1) and synthesizing SPR and AM (BSPS-ad2).
The MSE of these methods are 1.45 (BSPS-ad1) and 1.56 (BSPS-ad2).
Since BSPS-ad1 includes misspecified models, it would be natural that MSE is slightly inflated compared with the original BSPS. 
On the other hand, although BSPS-ad2 only combines models without using the true cutoff point, the inflation of prediction accuracy is quite limited.
This indicates the flexibility of prediction synthesis through BSPS.
In Supplementary Material~\ref{sec:add-4.1}, we provide detailed results on model weights as a function of bias and variance, estimated surface of the mean, and visualizing correlations among latent factors in the posterior distribution.

\subsection{Performance comparison}\label{sec:comp}
We next compare the performance of BSPS with other methods through Monte Carlo simulations. 
Let $s\in[0,1]^2$ be the spatial location generated from the uniform distribution of the region. 
The covariates $x_1\equiv x_1(s)$ and $x_2\equiv x_2(s)$ in the same way as in Section~\ref{sec:emp_behavior}.
Here we also generated $x_3, \ldots, x_p \ (p\geq 5)$ independently from $N(0,1)$.
We then consider the following two scenarios of data generating process:
\begin{align*}
\text{{\bf Scenario 1:}} \ \ \ y(s) &= w(s) + x_3^2 \exp(-0.3\|s\|^2) +s_2\sin(2x_2)+ \ep, \\
\text{{\bf Scenario 2:}} \ \ \ y(s) &=2w(s) +\frac12\sin(\pi x_1 x_2) + (x_3 - 0.5)^2 + \frac12x_4 + \frac14x_5 + \ep,  
\end{align*}
where $\ep\sim N(0, (0.7)^2)$ is an error term and $\omega(s)$ is an unobserved spatial effect following a zero-mean Gaussian process with covariance ${\rm Cov}(s, s') = (0.3)^2\exp(-\|s - s'\|/0.3)$. 
Note that the mean function in Scenario~1 changes according to location $s$ and the mean function in Scenario~2 is the well-known Friedman function \citep{friedman1991multivariate}.

We generated $300$ training samples and $100$ test samples of $(y,x,s)$ from each data generating process.
To predict test samples, we first consider the following prediction models: 
\begin{itemize}

\item[-]
GWR (geographically weighted regression): We used the R package ``spgwr" to fit the GWR model \citep{brunsdon1998geographically}, where the optimal bandwidth is selected by cross-validation. 

\item[-]
AM (additive model): We fitted the LAM model \citep{hastie1987generalized} with covariates $x_i$ by using the R package ``gam" with default settings for tuning parameters.

\item[-]
SPR (spatial regression): We fitted spatial linear regression with unobserved spatial effects modeled by 5-nearest neighbor Gaussian process, using R package ``spNNGP" \citep{spNNGP}, in which we draw 1000 posterior samples after discarding 1000 samples. 

\end{itemize}

We then synthesized the above three predictions.
We implemented BSPS with a 10-nearest neighbor Gaussian process and normal latent factors based on prediction values and their variances obtained from the three models.  
We generated 1000 posterior samples of the model coefficients, as well as the unknown parameters, after discarding the first 1000 samples as burn-in, to obtain the posterior predictive distribution of the test data. 
We also applied the variational Bayes BSPS (BSPS-VB), described in Section~\ref{sec:VB}, to obtain fast point predictions.
For comparison, we synthesized the three models via BMA and SA, as considered in Section~\ref{sec:emp_behavior}. 
Furthermore, we combine three models by a weighted average depending on prediction variance, as proposed by \cite{bates1969combination} (denoted by BG).
We also implement the super learner (SL) algorithm \citep[e.g.][]{davies2016optimal,van2007super}, where the optimal model weight is determined by the objective function based on K-fold cross validation.

As competitors of flexible spatial prediction methods, we adopted the following two models: 
\begin{itemize}
\item[-]
SRF (spatial random forest): We applied the recently proposed SRF \citep{saha2021random} using the R package ``RandomForestsGLS" with default settings (e.g. 50 trees). 

\item[-]
MGP (mixture of Gaussian process spatial regression): We applied mixtures of Gaussian process spatial regression, where each spatial regression is the same as SPR, and the number of components is set to 3. 
The posterior predictive distribution is obtained by generating 1000 posterior samples after discarding the first 1000 samples as burn-in.
\end{itemize}

\begin{figure}[!tb]
\centering
\includegraphics[width=14cm,clip]{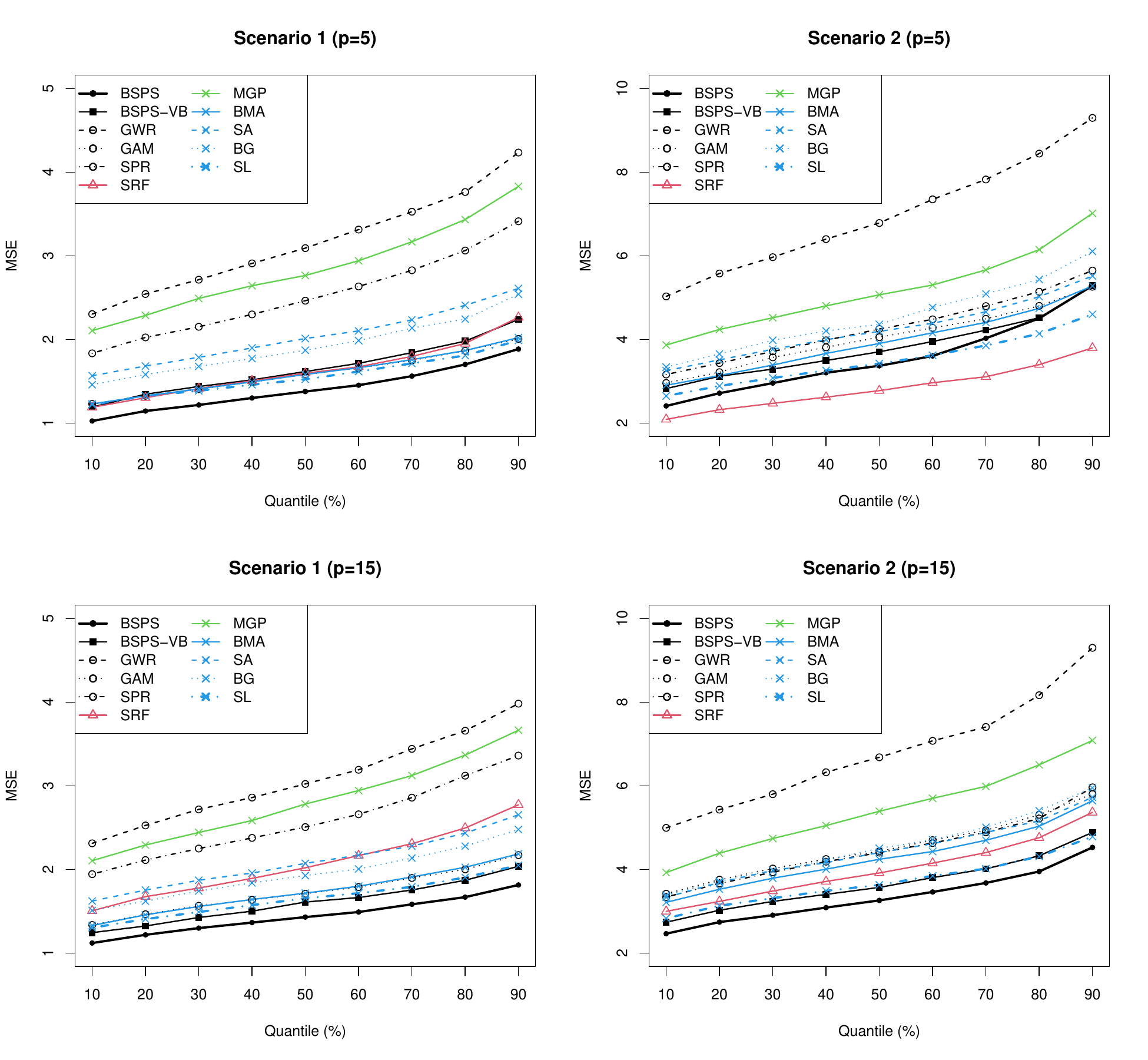}
\caption{Empirical 9 quantiles ($10\%, 20\%,\ldots,90\%$) of MSE values of 500 replications under two scenarios of data generating process with $p\in \{5, 15\}$. 
}
\label{fig:sim}
\end{figure}

To compare the performance in terms of point prediction, we computed the mean squared error (MSE) of the test data over 500 Monte Carlo replications, and present nine empirical quantiles ($10\%, 20\%,\ldots,90\%$) of MSE values in Figure~\ref{fig:sim}. 
BSPS provides the most accurate predictions in all the scenarios except for Scenario 2 with $p=5$, but the performance of BSPS is still the second best for most of the quantiles in this scenario. 
Notably, under a larger number of covariates ($p=15$), the performance of BSPS is considerably better than the other methods.
While BSPS can be seen as a mixture of Gaussian processes with random weights, the performance of MGP (a standard mixture approach using Gaussian processes) is not satisfactory, indicating that BSPS is not a mere mixture of Gaussian processes. 
It should be noted that the performance of BSPS tends to be superior to the other ensemble methods, BMA, SA, BG, and SL, which could be attributed to the data-dependent adaptation of the spatially varying model weight in BSPS.
Furthermore, BSPS improves the prediction accuracy of the three basic methods (GWR, AM, and SPR) in all scenarios, even though BMA and SA do not necessarily improve the performance, as confirmed in Scenario 1.  
The fast prediction method by BSPS-VB performs slightly worse than BSPS based on MCMC, though it still performs better than the standard ensemble methods, BMA and SA, in all scenarios.

We next evaluated the performance of $95\%$ interval prediction.
Here we focus on BSPS, GWR, AM, and SPR since SRF, BMA, and SA do not produce interval predictions.
The empirical coverage probability (CP) and average length (AL) under four scenarios are presented in Table \ref{tab:sim}.
While CPs of GWR and AM are not necessarily around the nominal level, CPs of BSPS and SPR are fairly close to the nominal level. 
Comparing BSPS and SPR, ALs of BSPS are considerably shorter than those of SPR, indicating both accuracy and efficiency of prediction intervals of BSPS.

\begin{table}
\caption{Coverage probability (CP) and average interval length (AL) of $95\%$ prediction intervals of test samples obtained from three basic models (GWR, AM, and SPR) and BSPS. 
The CP and AL are averaged over 500 replications of simulations.  
\label{tab:sim}}
\centering
\begin{tabular}{cccccccccccc}
\hline
&&& \multicolumn{4}{c}{CP ($\%$)} && \multicolumn{4}{c}{AL}\\
Scenario & $p$ &  & BSPS & GWR & AM & SPR &  & BSPS & GWR & AM & SPR \\
\hline
1 & 5 &  & 96.4 & 89.1 & 94.0 & 95.2 &  & 4.99 & 5.58 & 4.76 & 6.35 \\
1 & 15 &  & 95.9 & 90.8 & 92.3 & 95.5 &  & 4.90 & 5.81 & 4.67 & 6.47 \\
2 & 5 &  & 92.9 & 81.9 & 94.2 & 96.0 &  & 6.87 & 6.94 & 7.64 & 8.65 \\
2 & 15 &  & 93.8 & 86.1 & 92.2 & 96.0 &  & 6.91 & 7.69 & 7.55 & 8.81 \\
\hline
\end{tabular}
\end{table}

\section{Real Data Applications \label{sec:app}}
We consider two distinct real world applications to highlight the predictive performance of BSPS.
The first dataset is ecological: predicting the occurrence of Tsuga canadensis in Michigan, USA.
The second dataset is real estate: predicting apartment prices in Tokyo, Japan.
Both datasets are distinct, in that the ecological dataset is binary and deals with natural processes, while the real-estate dataset is continuous and deals with human economic activity.
This is done to illustrate the efficacy of BSPS and compare different methods under distinctly different situations, to provide a more holistic assessment.

\subsection{Occurrence of Tsuga canadensis in Michigan}\label{sec:app-binary}
The first real world application concerns the occurrence of Tsuga canadensis (Eastern hemlock) in Michigan, USA, analyzed in \cite{lany2020complementary}. 
The data comprise hemlock occurrence (binary outcome) on $17743$ forest stands across the state of Michigan. A set of covariates was also observed at each stand and can be used to explain the probability of hemlock occurrence. 
Covariates include minimum winter temperature, maximum summer temperature, total precipitation in the coldest quarter of the year, total precipitation in the warmest quarter of the year, annual actual evapotranspiration, and annual climatic water deficit. 
Spatial coordinates are recorded in longitude (lon) and latitude (lat).

There are several reasons why the prediction of the occurrence of Tsuga canadensis is relevant for this application.
As a long-lived, foundational species in Michigan, conservation is critical due to it being threatened by the hemlock woolly adelgid Adelges tsugae, an invasive sap-feeding insect.
Thus, predicting the occurrence is key in protecting the hemlock from this invasive species, by proactively making preventative measures.
Further, since the mechanism for hemlock habitat is not known, as hemlock does not occur in all suitable habitats, the interpretability of the prediction is relevant for future conservation.

To investigate the predictive performance of BSPS and compare it to the other methods, we randomly omitted 2000 spatial locations as the validation set, and used the remaining $n=15743$ samples as the training set. 
For the models to be synthesized in BSPS, we consider three Bernoulli models, $y_i\sim {\rm Ber}(e^{\psi_i}/(1+e^{\psi_i}))$, for $i=1,\ldots,n$, with the following specifications on the linear predictor, $p_i$, based on generalized linear models (GLM) and generalized additive models (GAM): 
\begin{align*}
\text{(GLM)} \ \ &\psi_i=\beta_0 + \sum_{k=1}^p \beta_k x_{ik},\\
\text{(GAM1)} \ \ &\psi_i=g_1({\rm lon}_i) + g_2({\rm lat}_i) + \sum_{k=1}^p \beta_k x_{ik},\\
\text{(GAM2)} \ \ &\psi_i=g_1({\rm lon}_i) + g_2({\rm lat}_i) + \sum_{k=1}^p f_k(x_{ik}),
\end{align*}
where $x_i=(x_{i1},\ldots,x_{ip})$ with $p=6$ is the vector of covariates described above, and $g_1, g_2$ and $f_k$ are unknown functions. 
We compute the occurrence probability in the validation dataset using the covariates and location information.
To synthesize these predictors through BSPS, we apply the logistic synthesis model (\ref{BPS-bin}) with $J=3$ latent factors corresponding to the above three predictors, and employed the nearest neighbor Gaussian process for the spatially varying model coefficients with $m=10$ nearest neighbors and an exponential covariance function. 
To construct distributions of the latent factors in BSPS, we used the Bayesian bootstrap \citep{rubin1981bayesian}, which fits models with randomly weighted observations, to extract uncertainty of estimation and prediction of $\psi_i$.
In this analysis, we used 100 bootstrap replications to compute the mean and variance of $\psi_i$ for each model.
We then generated 5000 posterior samples after discarding the first 2000 samples as burn-in, and generated random samples for the coefficient vectors in the validation set to compute predictions of binomial probability.
For comparison, we applied Bayesian model averaging (BMA), simple averaging (SA),  Bates-Granger averaging (BG), and super learner (SL) to combine the three models, as used in Section~\ref{sec:comp}.
Furthermore, we also applied a spatial logistic regression (SPR) model with spatial random effects modeled by the $m=10$ nearest neighbor Gaussian process, which was fitted using the R package ``spNNGP" with 5000 posterior samples after discarding the first 2000 samples. 
Spatial random forest \citep{saha2021random} was not considered for this application, since it does not support binary responses.

\begin{figure}[!tb]
\centering
\includegraphics[width=14cm,clip]{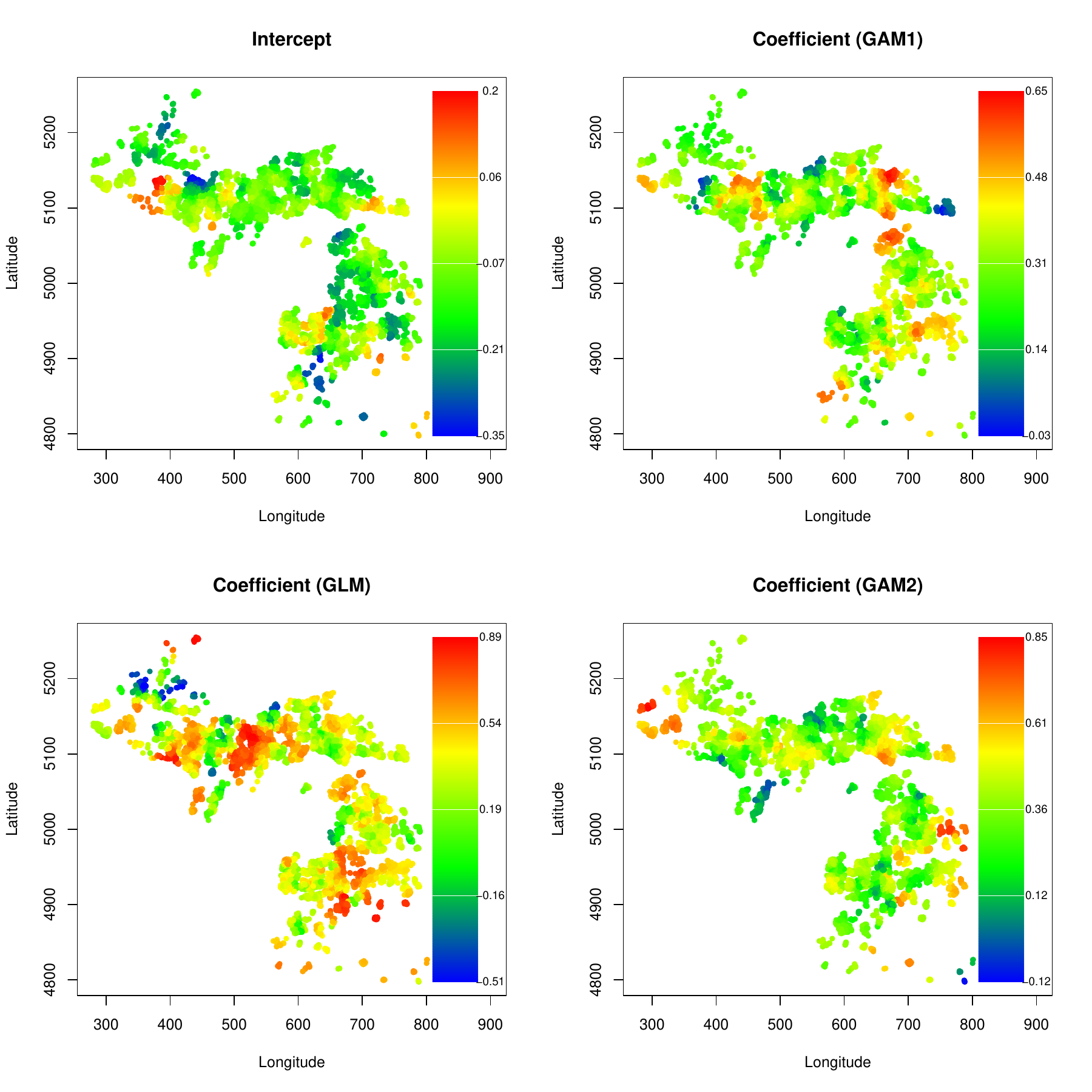}
\caption{Posterior means of spatially varying intercept and coefficients in the logistic BSPS model defined in (\ref{BPS-bin}).
\label{fig:bin-weight}
}
\end{figure}

\begin{figure}[!htb]
\centering
\includegraphics[width=14cm,clip]{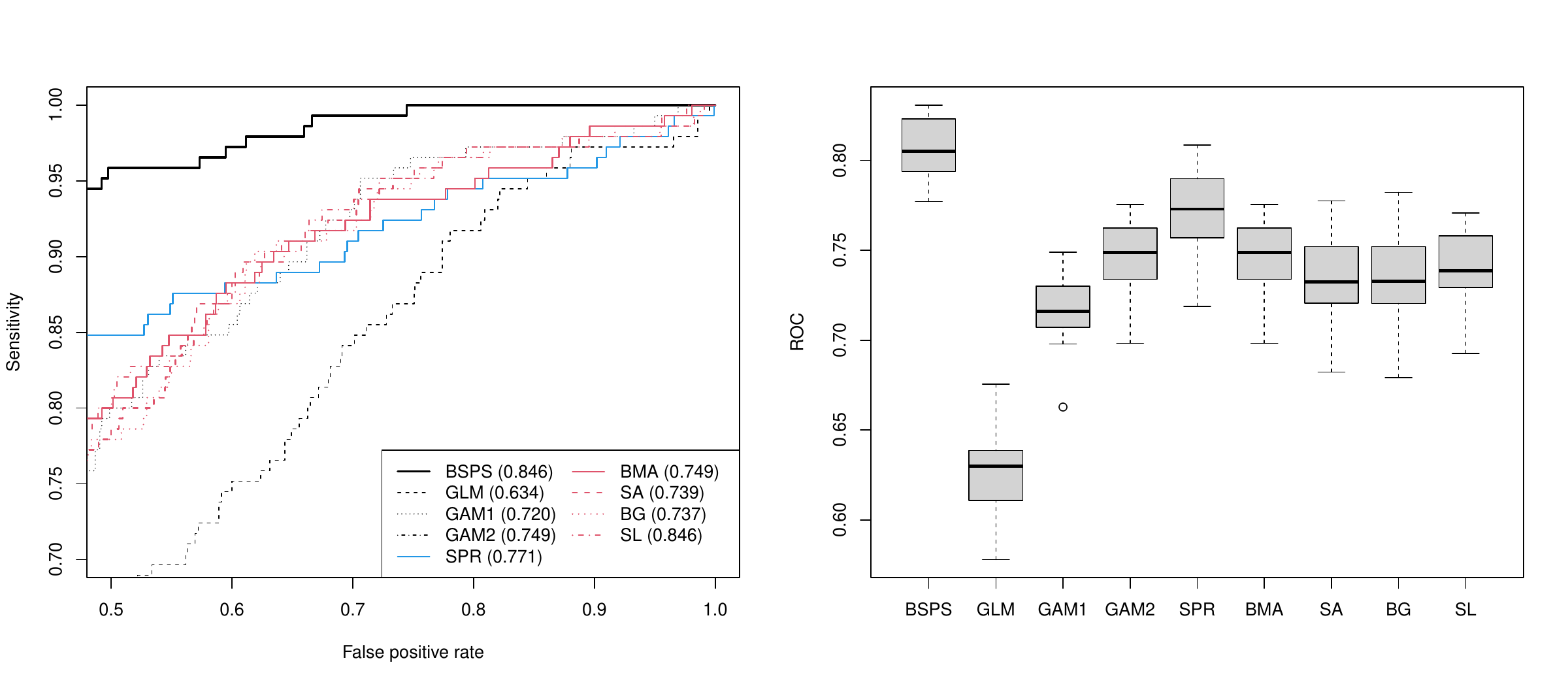}
\caption{Receiver operating characteristic (ROC) curves for various prediction methods (left), where the resulting values of area under the curve (AUC) are given in parenthesis. 
AUC values for the six methods under 20 replications (right).  
\label{fig:bin-ROC}
}
\end{figure}

In Figure~\ref{fig:bin-weight}, we present the spatial distributions of the posterior means of the spatially varying model coefficients, $\beta_j(s) \ (j=0,1,2,3)$, which shows how the importance of the three models change over regions. 
Particularly, it is interesting to see that the simplest GLM model is found to be more relevant for synthesis than the other models in some locations.  
This exemplifies how predictive performances vary spatially, where even simple models can be effective and relevant depending on the region. 
To compare the prediction performance in the test data, we compute the receiver operating characteristic (ROC) curves for the predicted binomial probabilities.
The results are presented in the left panel of Figure~\ref{fig:bin-ROC}, where the resulting values of area under the curve (AUC) are given in parenthesis.

We repeat the process, splitting the data and predicting the test data, 20 times, and report the boxplots of AUC values in the right panel of Figure~\ref{fig:bin-ROC}.
The figures show the superiority of BSPS to all other methods, including the existing model averaging methods, BMA, SA, and BG, in terms of AUC values.
An interesting phenomenon, though consistent across the studies in this paper, is that the AUC values of GLM, GAM1, and GAM2 are lower than that of SPR, but the AUC value of the synthesized prediction, through BSPS, is higher.
On the other hand, the AUC values of the ensemble methods, BMA, SA, BG, and SL, are at most the best performing model, GAM2. 
This indicates the effectiveness of BSPS in synthesizing simple models to provide more accurate predictions.

\subsection{Apartment prices in Tokyo}

Our second application is to apply BSPS to spatial predictions of apartment prices in the 23 wards in Tokyo, Japan. 
We used rent information using the ``Real Estate Data Library Door Data Nationwide 2013-2017 Data Set" (At Home Co., Ltd.) stored in the collaborative research system at the Center for Spatial Information Science, The University of Tokyo (\url{https://joras.csis.u-tokyo.ac.jp}).
The dataset contains the prices (yen), as well as auxiliary information on each room, for apartments handled by At Home, Inc. from 2013 to 2017. 
In this study, we used the samples collected in 2017, resulting in 22817 samples in total.
We adopted 11 covariates, five dummy variables of room arrangement, room area ($m^2$), balcony area ($m^2$), walking minutes from the nearest train station, age of building (month), indicator of newly-built room, and location floor.
For location information, the longitude and latitude information of each building, the name of the nearest train station, and the name of the ward are available. 
Since rooms in the same building share the same geographical information, we added a very small noise generated from $N_2(0, 10^{-3}I_2)$ to such rooms to avoid numerical instability. 
The room prices are log-transformed.

Similar to the ecological application in Section~\ref{sec:app-binary}, this application requires both accurate and interpretable predictions.
In terms of predictions, this is relevant for buyers, sellers, and real estate companies, but also for local governments to enact well-informed housing policies.
As apartment prices are not always reported, in the sense that they are not listed or prices are outdated, the prediction of these prices is crucial.
In terms of interpretation, one consideration that has received a lot of interest is the question of fairness and discrimination in these pricing models.
With the rise of black-box, machine learning models in real estate, the question of discriminatory pricing, which is not necessarily intended but happens due to the black-box nature of these algorithms, has been a major concern.
Having full interpretability, thus, is important for fair and transparent pricing practices.

We randomly omitted  2000 samples from the dataset, which are left as test samples. 
To construct the prediction models for room prices, we consider the following three types of models: 
\begin{itemize}
\item[-]
{\bf Station-level model:} \ \ The dataset is grouped according to 438 nearest train stations and simple linear regression with 5 covariates (walking minutes, room areas, and three dummy variables for room arrangement) is applied to each grouped sample.

\item[-]
{\bf Ward-level model:} \ \ The dataset is grouped according to the 23 wards and an additive model with 6 continuous covariates and three dummy variables for room arrangement is applied to each grouped sample.

\item[-]
{\bf Full model:} \ \  An additive model with 6 continuous covariates, two-dimensional location information, and five dummy variables for room arrangement.
\end{itemize}
Since the sample size that can be used to estimate the models increases in the order of station-level model, ward-level model, and full model, we vary the model complexity (e.g. number of parameters) in the three types of models. 
We also note that the three models are fully interpretable. 
The above models provide the means and variances for each training sample, and we synthesize the predictions with BSPS by assuming normality for each prediction model.  
With an exponential kernel in $m=10$ nearest neighbor Gaussian process for spatially varying model coefficients, we generated 7000 posterior samples after discarding 3000 samples as burn-in. 
For comparison, we applied spatial random forest (SRF) with 100 trees and spatial regression (SPR), as used in Section \ref{sec:comp}, to predict the room prices in the test data. 
Furthermore, we fitted the extreme gradient boosting tree \citep[XGB;][]{chen2016xgboost} with 1000 trees using the R package ``xgboost", where the optimal number of trees was selected via 5-fold cross validation and learning and sub-sampling rates set to $0.01$ and $0.1$, respectively. 
Note that the 11 covariates and two-dimensional location information are used to estimate SRF, SPR, and XGB, as used in the three levels of models in BSPS.

\begin{figure}[!tb]
\centering
\includegraphics[width=14cm,clip]{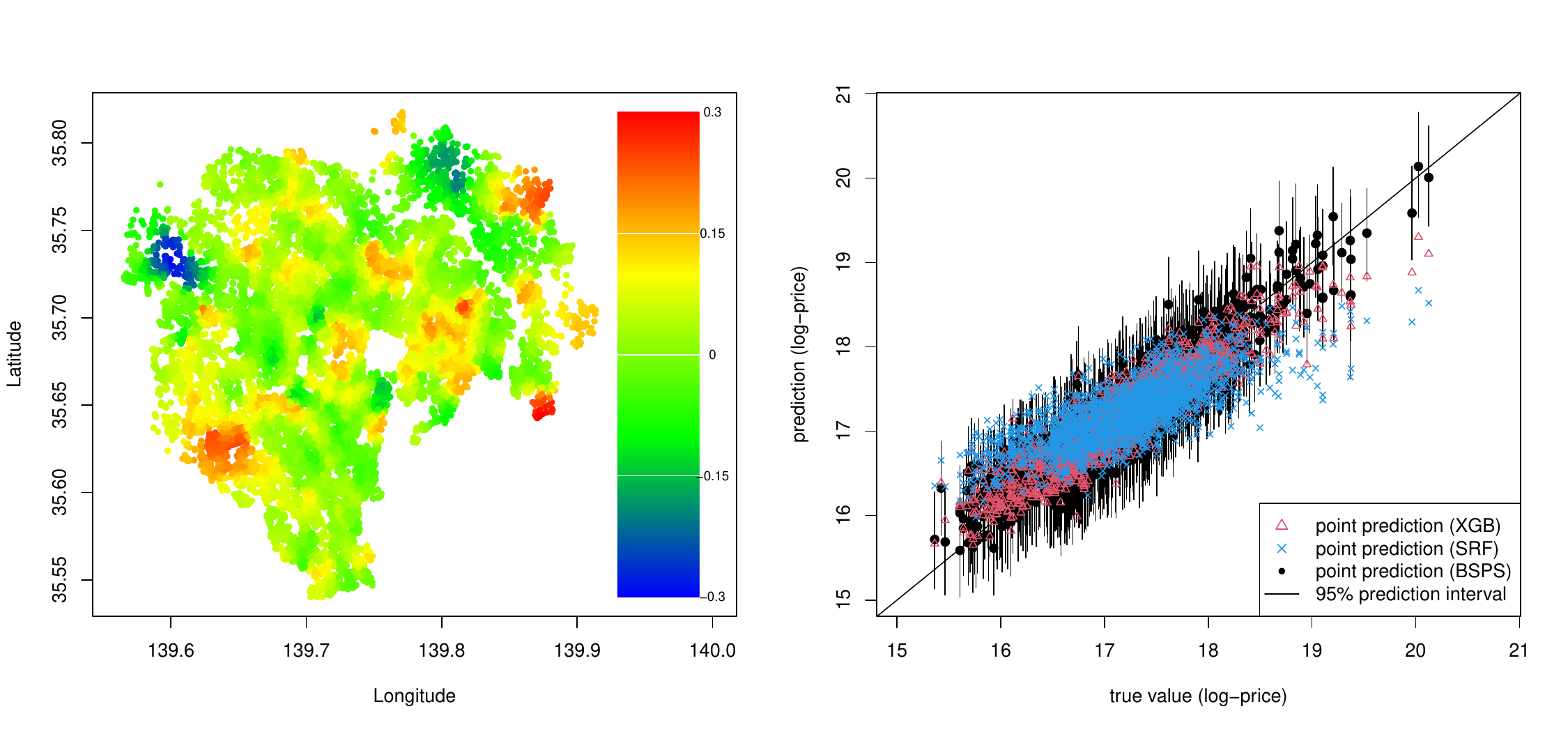}
\caption{Left: Spatial plot of posterior means of the intercept term. Right: Point predictions of XGB (red circle), SRF (blue circle), and BSPS (black circle) with the $95\%$ prediction intervals (vertical lines) based on the predictive distributions given by BSPS.  
\label{fig:mansion}
}
\end{figure}

The left panel in Figure \ref{fig:mansion} reports the spatial plot of $\beta_0(s_i)$, namely, the intercept term of BSPS.
Since the intercept term captures the variability not captured by the model set, it effectively represents the model set uncertainty.
Looking at the figure, we can see that the intercept is the largest in absolute value in certain regions.
Each of these regions has different reasons for why the model set uncertainty is so high, some are due to new development skewing prices, some are due to heterogeneity in popular residential areas, and some are due to changes in disclosure rules.
While the reason varies, the output of BSPS gives a clear and transparent indication for further inquiry.

We now consider comparing the predictive accuracy of each method for this application.
The MSE values for predicting the test samples are 
$$
\text{BSPS}: 0.241, \ \ \ \text{XGB}: 0.257, \ \ \ 
\text{SPR}: 0.268,  \ \ \ \text{SRF}: 0.425, 
$$
where, again, BSPS provides superior prediction accuracy.
As with the previous applications, it should be noted that the resulting predictors made by BSPS are interpretable, while XGB and SRF are not. 
We computed the $95\%$ prediction intervals from the posterior distributions for the test samples.
In the right panel in Figure \ref{fig:mansion}, we report the point predictions of XGB, SRF, and the $95\%$ prediction intervals for BSPS.
First, the relatively large MSE values of SRF come from the degeneracy of the point prediction, that is, the point prediction is much less variable than the true prices. 
This can be seen by the fact that the predictions are mostly horizontal, not deviating much from the mean.
While XGB provides reasonable point prediction overall, XGB tends to under-predict the large true price, as with the ecological application. 
On the other hand, BSPS provides accurate point predictions and $95\%$ prediction intervals with reasonable interval lengths regardless of the true price. 
The coverage proportion is $97.3\%$, which is well-calibrated for these tasks.

\section{Concluding Remarks}
\label{sec:conc}

Bayesian predictive synthesis provides a theoretically and conceptually sound framework to synthesize density prediction.
Utilizing this framework, we develop a spatially varying synthesis method for the context of spatial data. 
With this new method, we can dynamically calibrate, learn, and update coefficients as the data changes across a spatial region. 
The simulations and real world applications demonstrate the efficacy of BSPS compared to conventional spatial models and modern machine learning techniques.  
Specifically, by dynamically synthesizing the predictive distribution from the models, BSPS can improve point and distributional predictions.
Additionally, posterior inference on the full spatial region gives the decision maker information on how each model is related, and how their relationship changes across a region. 
In addition to the applications in this paper, our proposed framework can be applied to other fields, including, but not limited to, weather, GPS systems, and sports player tracking data. Further studies exploring different uses of BSPS, as well as specific developments catered towards a specific dataset, are of interest.

Regarding scalable computation algorithms for BSPS, it may be possible to use other types of scalable Gaussian processes, such as predictive process \citep{Banerjee2008}, meshed Gaussian process \citep{peruzzi2020highly}, and fused Gaussian process \citep{ma2020fused}.
We leave the potential use of these techniques as future work. 
Apart from MCMC-based algorithms, the integrated nested Laplace approximation \citep{rue2009approximate} may be an appealing strategy for fast computation. 
However, the latent factor spatially varying coefficient model (\ref{SVC2}) has $2(J+1)$ hyperparameters, which limits the use of the integrated nested Laplace approximation when $J$ is not small (e.g. $J\geq 3$).

In applying BSPS, one essential assumption is that we have prediction uncertainty of models to be synthesized. 
However, some models (especially machine learning methods) provide only point prediction, and extracting the uncertainty of prediction may not be straightforward.
A possible remedy is to use the Bayesian bootstrap \citep{rubin1981bayesian} to roughly capture the prediction uncertainty, as used in Section~\ref{sec:app-binary}.
Although such bootstrap-based uncertainty quantification works numerically well in our application,  detailed discussions, including theoretical arguments, are left for future works.

Finally, there are several ways to extend or apply the current BSPS approach. 
The first is to extend BSPS to spatio-temporal or multivariate data.
This can potentially be done by using the recently developed techniques of graphical Gaussian processes \citep{dey2020graphical,peruzzi2022spatial}.
Moreover, BSPS can be used, not only for synthesizing multiple models, but also for saving computational cost under a large number of covariates. 
Specifically, if the number of covariates, say $p$, is very large, the standard spatially varying coefficient model \citep{gelfand2003spatial} requires $p+1$ Gaussian processes for modeling spatially varying coefficients, which is computationally burdensome. 
On the other hand, it would be possible to first apply multiple, say $J$, regression models to get univariate spatial predictors, and then combine the multiple prediction models via BSPS. 
This reduces the number of Gaussian processes from $p+1$ to $J+1$, which makes the computation much less burdensome.

\section*{Acknowledgement}
This work is partially supported by Japan Society for Promotion of Science (KAKENHI) grant numbers 21H00699.

\bibliographystyle{chicago}
\bibliography{ref}

\newpage
\setcounter{page}{1}
\setcounter{equation}{0}
\setcounter{section}{0}
\setcounter{figure}{0}
\setcounter{table}{0}
\setcounter{thm}{0}
\renewcommand{\thesection}{S\arabic{section}}
\renewcommand{\theequation}{S\arabic{equation}}
\renewcommand{\thefigure}{S\arabic{figure}}
\renewcommand{\thetable}{S\arabic{table}}
\renewcommand{\thethm}{S\arabic{thm}}

\vspace{1cm}
\begin{center}
{\LARGE
{\bf Supplementary Material for \\
``Bayesian Spatial Predictive Synthesis"}
}
\end{center}

\vspace{0.8cm}
This Supplementary Material provides the following.
Details of the sampling algorithm under nearest neighbor Gaussian processes are in Section~\ref{sec:NNGP-pos}.
The derivation of the mean-field variational Bayes algorithm (Algorithm 1 in the main document) is in Section~\ref{sec:VB-derivation}. 
Section~\ref{sec:par-lf} describes the implementation, computation, and numerical simulation of particle latent factors (i.e. when the predictive distribution is given as posterior samples).
Finally, Section~\ref{sec:add-4.1} describes the two alternative latent factor specifications (spatially correlated latent factors and overlap corrected latent factors), and their numerical comparison.
The exact minimaxity result, its proof, and an overview of Kiefer's theorem in Section~\ref{sec:minimax}.

\section{Sampling steps under nearest neighbor Gaussian process}
\label{sec:NNGP-pos}

The use of the $m$-nearest neighbor Gaussian process for $\beta_j(s)$ leads to a multivariate normal distribution with a sparse precision matrix for $\beta_j(s_1),\ldots,\beta_j(s_n)$, defined as 
$$
\pi(\beta_j(s_1),\ldots,\beta_j(s_n))
=\prod_{i=1}^n \phi(\beta_j(s_i);\B_j(s_i)\beta_j(N(s_i)), \tau_j \F_j(s_i)), \ \ \ \ j=0,\ldots,J
$$
where 
\begin{align*}
\B_j(s_i)&=C_j(s_i, N(s_i))C_j(N(s_i), N(s_i))^{-1}, \\ 
\F_j(s_i)&=C_j(s_i,s_i)-C_j(s_i, N(s_i))C_j(N(s_i), N(s_i))^{-1}C_j(N(s_i), s_i),
\end{align*}
and $N(s_i)$ denotes an index set of $m$-nearest neighbors of $s_i$.
Here $C_j(\cdot,\cdot)$ is the same correlation function used in the original Gaussian process for $\beta_j(s)$.

The full conditional distributions of the latent factors, $f_{ji}$, and error variance, $\sigma^2$, are the same as the ones given in the main document. 
The full conditional distributions of the other parameters are given as follows: 

\begin{itemize}
\item[-]
(Sampling of spatially varying weights) \ \ For $i=1,\ldots,n$, the full conditional distribution of $(\beta_0(s_i),\ldots,\beta_J(s_i))$ is given by $N(A_{i}^{(\beta)}B_{i}^{(\beta)}, A_{i}^{(\beta)})$, where 
\begin{align*}
&A_{i}^{(\beta)}=\left\{\frac{f_if_i^{\top}}{\sigma^2}+{\rm diag}(\gamma_{0i},\ldots,\gamma_{Ji})\right\}^{-1}, \ \ \ \  \gamma_{ji}=\frac{1}{\tau_j F_j(s_i)}+\sum_{t;s_i\in N(t) }\frac{B_j(t; s_i)^2}{\tau_j F_j(t)}, \\
&B_{i}^{(\beta)}=\frac{f_iy_i}{\sigma^2}+(m_{0i},\ldots,m_{Ji})^\top, \\ &m_{ji}=\frac{B_j(s_i)^\top \beta_j(N(s_i))}{\tau_jF_j(s_i)} +\sum_{t;s_i\in N(t) }\frac{B_j(t;s_i)}{\tau_jF_j(t)}\Big\{\beta_j(t)-\sum_{s\in N(t), s\neq s_i}B_j(t;s) \beta_j(s)\Big\}
\end{align*}
where $f_i=(1, f_{1i},\ldots,f_{Ji})$ and $B_j(t;s)$ denotes the scalar coefficient for $\beta_j(s_i)$ among the element of the coefficient vector $B_j(t)$.

\item[-]
(Sampling of $\tau_j$) \ \ 
For $j=0,\ldots,J$, the full conditional distribution of $\tau_j$ is 
$$
{\rm IG}\left(a_\tau+\frac{n}{2}, b_{\tau}+\frac12\sum_{i=1}^n \frac{\big\{\beta_j(s_i)-B_j(s_i)\beta_j(N(s_i))\big\}^2}{F_j(s_i)}\right).
$$

\item[-]
(Sampling of $g_j$) \ \ 
For $j=0,\ldots,J$, the full conditional distribution of $g_j$ is proportional to 
$$
\prod_{i=1}^n \phi(\beta_j(s_i);B_j(s_i; \theta_j)\beta_j(N(s_i)), \tau_jF_j(s_i;g_j)), \ \ \  g_j\in (\underline{g}, \overline{g}),
$$
where 
\begin{align*}
B_j(s_i;g_j)&=C_j(s_i, N(s_i);g_j)C_j(N(s_i), N(s_i);g_j)^{-1}, \\ 
F_j(s_i;g_j)&=C_j(s_i,s_i;g_j)-C_j(s_i, N(s_i);g_j)C_j(N(s_i), N(s_i);g_j)^{-1}C_j(N(s_i), s_i;g_j),
\end{align*}
and $C_j(\cdot, \cdot;g_j)$ is the correlation function with spatial range $g_j$.
\end{itemize}

\section{Derivation of variational Bayes algorithm}
\label{sec:VB-derivation}

Remember that the mean filed variational Bayes (MFVB) approximates the posterior distributions through the form
$$
q(\{f_{ji}\}, \{\beta_j\}, \{\tau_j\}, \{g_j\},\sigma^2)=q(\sigma^2)\prod_{j=0}^J q(\beta_j)q(\tau_j)q(g_j)\prod_{i=1}^n q(f_{ji}).
$$
It is known that the optimal form of the variational posterior is given by, for example, $q(\beta_j)\propto \exp(\mathbb{E}_{-\beta_j}[\log p(y,\Theta)])$, where $\Theta=(\{f_{ji}\}, \{\beta_j\}, \{\tau_j\}, \{g_j\},\sigma^2)$ and $\mathbb{E}_{-\beta_j}$ denotes the expectation with respect to the marginal variational posterior of the parameters other than $\beta_j$.
From the forms of full conditional posterior distributions given in the main document, we can use the following distributions as optimal distributions: 
\begin{align*}
&q(\sigma^2)\sim {\rm IG}(\at_{\sigma}, \bt_{\sigma}), \ \ \ \
q(\tau_j)\sim {\rm IG}(\at_{\tau_j}, \bt_{\tau_j}), \ \ \ \
q(\beta_j)\sim N(\mut_{j}, \Sit_{j}), \\
&q(g_j)\sim {\rm D}(\pt_{j1},\ldots,\pt_{jL}), \ \ \ \ 
q(f_{ji}) \sim N(\mt_{ji}, \st^2_{ji}),
\end{align*}
where ${\rm D}(\pt_{j1},\ldots,\pt_{jL})$ is a discrete distribution on $\{\eta_1,\ldots,\eta_L\}$ such that $P(g_j=\eta_\ell)=\pt_{j\ell}$.
The derivation of the updating steps of MFVB is given as follows.

\begin{itemize}
\item[-] {\bf (update of $f_{ji}$)} \ \ It follows that 
$$
\mathbb{E}_{-f_{ji}}[\log p(y,\Theta)]
=(\text{const.})-\frac12 f_{ij}^2(A_{ji}^f)^{-1} +f_{ji}B_{ji}^{f}, 
$$
where
\begin{align*}
A_{ji}^f
&=\left(\frac1{b_{ji}}+\mathbb{E}_q[\beta_{ji}^2]\mathbb{E}_q\left[\frac{1}{\sigma^2}\right]\right)^{-1}
=\left\{\frac1{b_{ji}}+(\mut_{ji}^2+\Sit_{jii})\frac{\at_{\sigma}}{\bt_{\sigma}}\right\}^{-1},
\end{align*}
and
\begin{align*}
B_{ji}^f
&=\frac{a_{ji}}{b_{ji}}+\mathbb{E}_q[\beta_{ji}]\mathbb{E}_q\left[\frac{1}{\sigma^2}\right]\bigg(y_i-\mathbb{E}_q[\beta_{0i}]-\sum_{k\neq j}\mathbb{E}_q[\beta_{ki}]\mathbb{E}_q[f_{ki}]\bigg)\\
&=\frac{a_{ji}}{b_{ji}}+\mut_{ji}\frac{\at_{\sigma}}{\bt_{\sigma}}\bigg(y_i-\mut_{0i}-\sum_{k\neq j}\mut_{ki}\mt_{ki}\bigg).
\end{align*}
Then, the parameters in the variational posterior of $f_{ji}$ can be updated as $\mt_{ji}=A_{ji}^fB_{ji}^f$ and $\st_{ji}^2=A_{ji}^f$.

\item[-] {\bf (update of $\beta_j$)} \ \ 
It follows that 
$$
\mathbb{E}_{-\beta_j}[\log p(y,\Theta)]
=(\text{const.})-\frac12\beta^{\top}(A_j^{(\beta)})^{-1}\beta +\beta^\top B_j^{(\beta)},
$$
where
\begin{align*}
A_j^{(\beta)}
&=\left\{\mathbb{E}_q[\Omega_j]\mathbb{E}_q\left[\frac1{\sigma^2}\right] + \mathbb{E}_q[H(g_j)^{-1}]\mathbb{E}_q\left[\frac1{\tau_j}\right]\right\}^{-1}
=\left\{\Omega^{\ast}_j\frac{\at_{\sigma}}{\bt_{\sigma}} + \sum_{\ell=1}^L \pt_{j\ell}H(\eta_\ell)^{-1}\frac{\at_{\tau_j}}{\bt_{\tau_j}}\right\}^{-1} \\
B_j^{(\beta)}
&=\mathbb{E}_q\left[\frac1{\sigma^2}\right] \mathbb{E}[F_j]\circ
\bigg(y-\mathbb{E}_q[\beta_0]-\sum_{k\neq j}\mathbb{E}_q[\beta_k]\circ \mathbb{E}_q[F_k]\bigg)\\
&=\frac{\at_{\sigma}}{\bt_{\sigma}} \mt_j\circ
\bigg(y-\mut_0-\sum_{k\neq j}\mut_k\circ \mt_k\bigg),
\end{align*}
where $\Omega^{\ast}={\rm diag}(\mt_{j1}^2+\st_{j1}^2,\ldots,\mt^2_{jn}+\st^2_{jn})$.
Then, the parameters in the variational posterior of $\beta_j$ can be updated as $\mut_{j}=A_j^{(\beta)}B_j^{(\beta)}$ and $\Sit_{j}=A_j^{(\beta)}$.

\item[-] {\bf (update of $\tau_j$)}  \ \ 
It follows that 
\begin{align*}
\mathbb{E}_{-\tau_j}[\log p(y,\Theta)]
&=(\text{const.}) -\left(\frac{n}{2}+a_{\tau}+1\right)\log \tau_j -\frac1{\tau_j}\left(b_\tau + \frac12\mathbb{E}_q\left[\beta_j^\top H(g_j)^{-1}\beta_j\right]\right),
\end{align*}
noting that 
$$
\mathbb{E}_q\left[\beta_j^\top H(g_j)^{-1}\beta_j\right]
={\rm tr}\left\{\mathbb{E}_q[\beta_j\beta_j^\top]\mathbb{E}_q[H(g_j)^{-1}]\right\}
={\rm tr}\left\{(\mut_j\mut_j^\top + \Sit_j)\sum_{\ell=1}^L\pt_{j\ell}H(\eta_\ell)^{-1}\right\}.
$$
Then, the parameters in the variational posterior of $\tau_j$ can be updated as 
$$
\at_{\tau_j}=a_{\tau}+\frac{n}{2}, \ \ \ \ \bt_{\tau_j}=b_\tau + \frac12{\rm tr}\left\{(\mut_j\mut_j^\top + \Sit_j)\sum_{\ell=1}^L\pt_{j\ell}H(\eta_\ell)^{-1}\right\}.
$$

\item[-] {\bf (update of $g_j$)}  \ \ It follows that  
$$
\mathbb{E}_{-g_j}[\log p(y,\Theta)]
=(\text{const.}) -\frac12\log |H(g_j)| -\frac12\mathbb{E}_q\left[\frac1{\tau_j}\right]\mathbb{E}_q[\beta_j^\top H(g_j)^{-1}\beta_j], 
$$
where $\mathbb{E}_q[\beta_j^\top H(g_j)^{-1}\beta_j]={\rm tr}\{(\mut_j\mut_j^\top + \Sit_j)H(g_j)^{-1}\}$.
Then, the parameters in the variational posterior of $g_j$ can be updated as 
$$
\pt_{j\ell}
=\frac{|H(\eta_\ell)|^{-1/2}\exp\left(-\at_{\tau_j}{\rm tr}\{(\mut_j\mut_j^\top + \Sit_j)H(\eta_\ell)^{-1}\}/2\bt_{\tau_j}\right)}
{\sum_{\ell'=1}^L|H(\eta_{\ell'})|^{-1/2}\exp\left(-\at_{\tau_j}{\rm tr}\{(\mut_j\mut_j^\top + \Sit_j)H(\eta_{\ell'})^{-1}\}/2\bt_{\tau_j}\right)}.
$$

\item[-] {\bf (update of $\sigma^2$)}  \ \ It follows that  
$$
\mathbb{E}_{-\sigma^2}[\log p(y,\Theta)]
=(\text{const.}) -\frac{n}{2}\log\sigma^2 -\frac1{2\sigma^2}\sum_{i=1}^n\mathbb{E}_q\bigg[\bigg(y_i-\beta_{0i}-\sum_{j=1}^J\beta_{ji}f_{ji}\bigg)^2\bigg],
$$
where 
\begin{align*}
I_q(\sigma^2) &\equiv \sum_{i=1}^n  \mathbb{E}_q\bigg[\bigg(y_i-\beta_{0i}-\sum_{j=1}^J\beta_{ji}f_{ji}\bigg)^2\bigg]\\
&=
\bigg(y-\mut_0-\sum_{j=1}^J\mut_j\circ \mt_j\bigg)^\top \bigg(y-\mut_0-\sum_{j=1}^J\mut_j\circ \mt_j\bigg)
+
{\rm tr}(\Sit_0) \\
& \ \ \ + \sum_{j=1}^J {\rm tr}\left\{(\mut_j\mut_j^\top  + \Sit_j)\circ (\mt_j\mt_j^\top+\St_j)\right\} - \sum_{i=1}^n\sum_{j=1}^J \mt_{ji}^2\mut_{ji}^2
\end{align*}
with $\St_j={\rm diag}(\st_{j1}^2,\ldots,\st_{jn}^2)$.
Then, the parameters in the variational posterior of $\sigma^2$ can be updated as 
$$
\at_{\sigma}=a_{\sigma}+\frac{n}{2}, \ \ \ \ \bt_{\sigma}=b_{\sigma} + \frac12I_q(\sigma^2).
$$

\end{itemize}

\section{Particle latent factors}
\label{sec:par-lf}

\subsection{Posterior computation under particle latent factors}
Suppose that the distribution of the latent factor $f_{ji}$ is a discrete uniform distribution on $\{c_{ji}^{(1)},\ldots,c_{ji}^{(P)}\}$, where $P$ is the number of particles. 
These particles may be the posterior predictive distribution of some Bayesian models. 
Under the discrete distribution of $f_{ji}$, the posterior computation algorithm of BSPS will change only for $f_{ji}$.
Given the other parameters and latent variables, the full conditional distribution of $f_{ji}$ is a discrete distribution with the probability that $f_{ji}=c_{ji}^{(p)}$ being 
$$
\frac{\phi\Big(y_i; \beta_{0i}+\beta_{ji}c_{ji}^{(p)}+\sum_{k\neq j}^J\beta_{ki}f_{ki}, \sigma^2\Big)}{\sum_{p'=1}^P\phi\Big(y_i; \beta_{0i}+\beta_{ji}c_{ji}^{(p)}+\sum_{k\neq j}^J\beta_{ki}f_{ki}, \sigma^2\Big)}, \ \ \ \ p=1,\ldots,P.
$$
Then, $f_{ji}$ can be generated from the above multinomial distribution.

\subsection{Comparison of Gaussian and particle latent factors}
Using the same simulation scenarios in Section~4.2 in the main text, we compared the particle latent factors and Gaussian approximated latent factors as used in the default BSPS.
As in Section~4.2, we adopted three models, GWR, AM, and SPR, and the pseudo-posterior samples of predictive distributions of GWR and AM are drawn by using the Bayesian bootstrap \citep[e.g.][]{rubin1981bayesian}. 
We generated 500 pseudo-posterior samples of GWR and AM as well as 500 posterior samples of SPR after discarding 1000 samples, and used the discrete distribution as the particle latent factors in BSPS.
For comparison, we computed the means and variances of 500 (pseudo-)posterior samples and adopted Gaussian latent factors having the means and variances. 
For two types of BSPS, 1000 posterior samples of posterior predictive distributions of test data are generated after discarding the first 1000 samples. We then compute posterior means and $95\%$ credible intervals, and evaluate the performance by MSE, CP, and AL, as used in Section~4.2.

In Table~\ref{tab:sim-supp-particle}, we report the three performance measures averaged over 200 Monte Carlo replications. 
The results show that the difference between using particle and Gaussian latent factors is quite limited in all the scenarios.
Since the particle latent factors can be computationally intensive when the number of particles is large, the use of approximated Gaussian distribution would be practically useful.

\begin{table}[!htb]
\caption{Mean squared errors (MSE) of point prediction, Coverage probability (CP), and average interval length (AL) of $95\%$ prediction intervals of test samples obtained by particle and Gaussian (default) latent factors.
The reported MSE, CP, and AL are averaged over 200 replications of simulations. 
\label{tab:sim-supp-particle}}
\centering
\medskip
\begin{tabular}{cccccccccccccccc}
\hline
&&& \multicolumn{2}{c}{MSE}&&\multicolumn{2}{c}{CP (\%)}&&\multicolumn{2}{c}{AL}\\
Scenario & $p$ &  & Default & Particle &  & Default & Particle &  & Default & Particle \\
\hline
1 & 5 &  & 1.34 & 1.35 &  & 97.3 & 97.2 &  & 5.02 & 5.02 \\
1 & 15 &  & 1.41 & 1.41 &  & 96.0 & 96.1 &  & 4.87 & 4.87 \\
2 & 5 &  & 3.20 & 3.24 &  & 95.9 & 96.0 &  & 7.21 & 7.21 \\
2 & 15 &  & 3.24 & 3.25 &  & 95.2 & 95.1 &  & 7.12 & 7.13 \\
\hline
\end{tabular}
\end{table}

\section{Additional Results in Section~4.1}
\label{sec:add-4.1}

Here, we show additional numerical results in Section~4.1 in the main text. 
In particular, we present posterior correlation among different latent factors and demonstrate how the bias and variance of each model are related to the model weight in BSPS.

\subsection{Correlation in the posterior distribution of the latent factors}

\begin{figure}[!htb]
\centering
\includegraphics[width=14cm,clip]{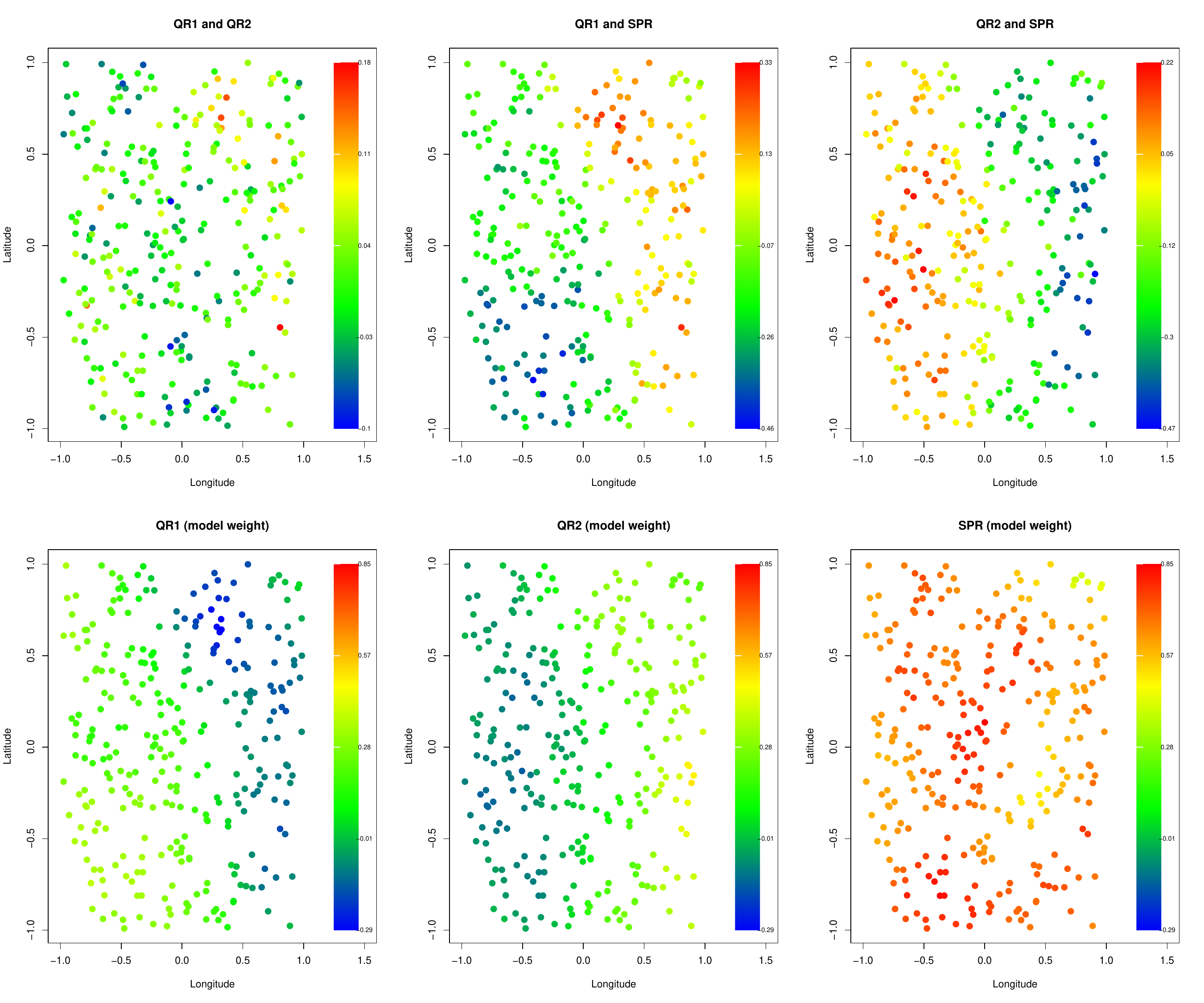}
\caption{Spatial distribution of pair-wise correlations between posterior distributions of different latent factors (upper) and estimated model weights (lower).  
\label{fig:correlation}
}
\end{figure}

In the proposed BSPS, the prior distribution of latent factors is mutually independent, but their posterior can be correlated. 
To demonstrate this issue, we calculated pair-wise posterior correlations among three models (QR1, QR2, and APR) at each location.
In Figure~\ref{fig:correlation}, we show posterior correlation with estimated model weights.
Since QR1 and QR2 are learned in disjoint regions, it would be natural to see that the posterior correlations are low overall.
On the other hand, posterior of QR1 and SPR are correlated in some regions while model weights for QR1 are relatively small in the region with a large correlation between QR1 and SPR. 
Similar results can be confirmed for QR2 and SPR.

\subsection{Connection of model weight to bias and variance}

\begin{figure}[!htb]
\centering
\includegraphics[width=14cm,clip]{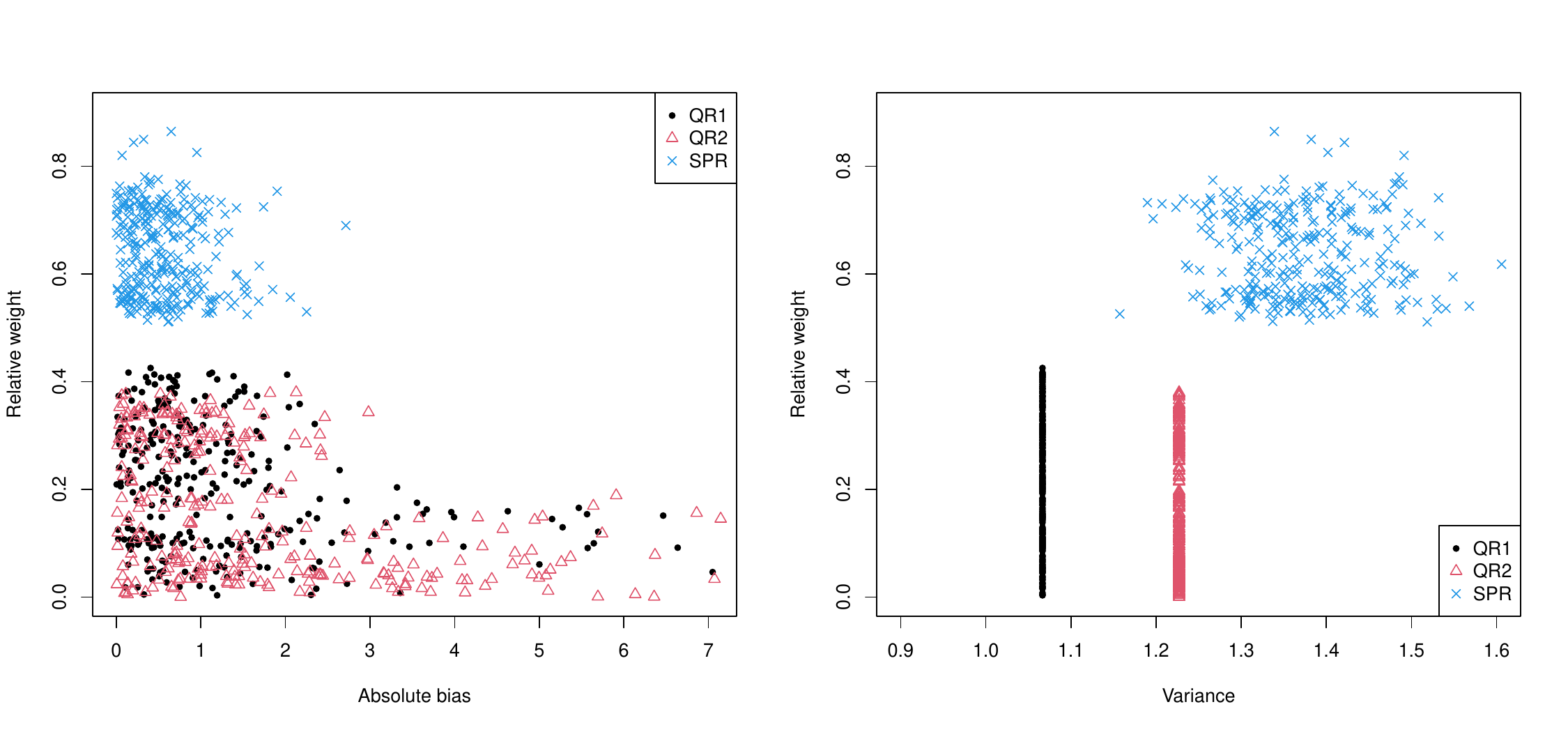}
\caption{Spatially varying model weight as a function of bias and variance. 
\label{fig:bias-variance}
}
\end{figure}

In Figure~\ref{fig:bias-variance}, we show the estimated model weights against absolute bias and variance of the three models. 
First, it can be seen that the model weight does not have clear relationships with variance unlike the Bates-Granger averaging used in Section~4.2. 
Secondly, the model weights tend to be large when the absolute bias is small.

\subsection{Estimated surface of response variable}

\begin{figure}[!htb]
\centering
\includegraphics[width=14cm,clip]{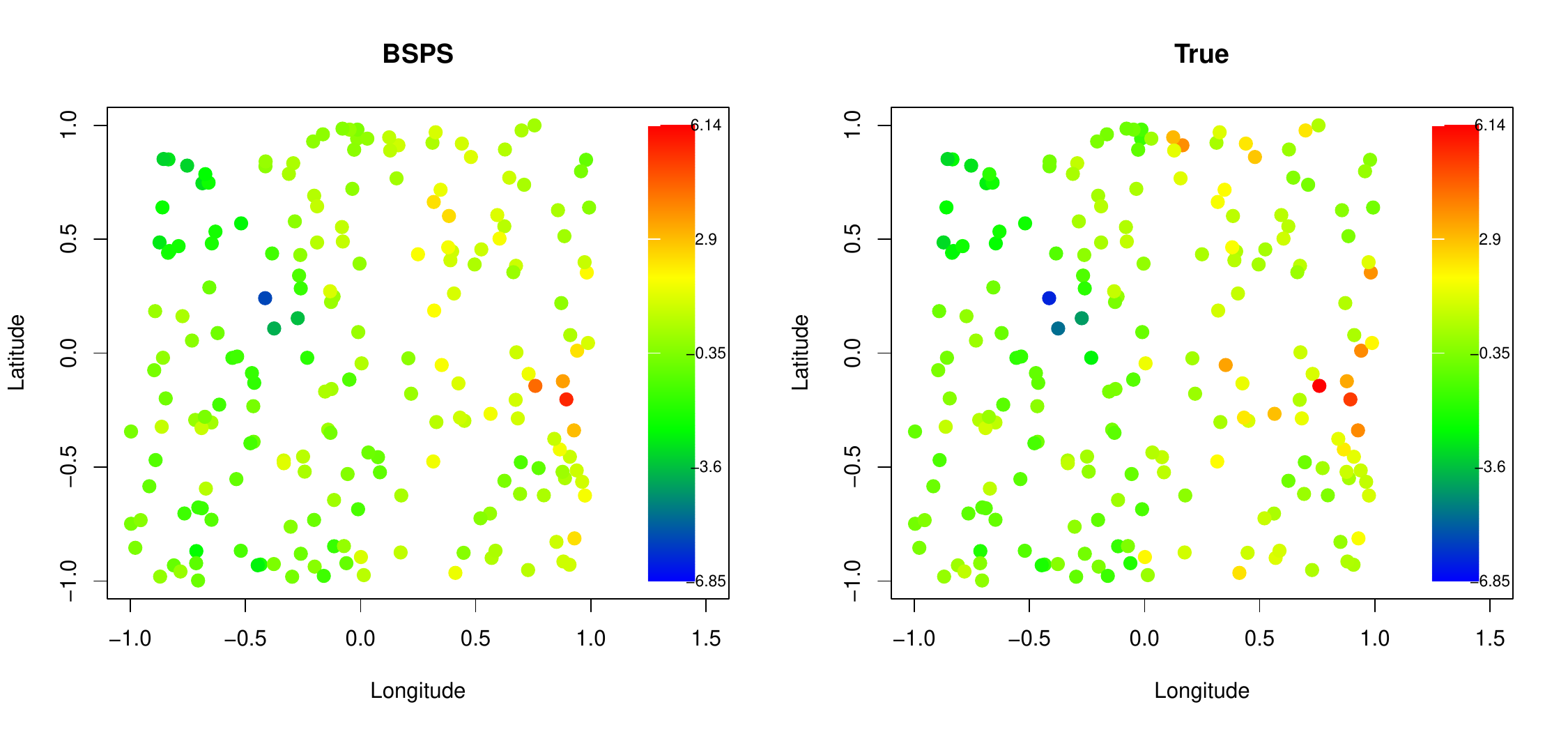}
\caption{Estimated surface of the response variable obtained by BSPS-ad1 (left) and its true surface (right).
\label{fig:surface}
}
\end{figure}

In Figure~\ref{fig:surface}, we show the point prediction of test data made by BSPS with the true test values. 
It shows that the prediction surface is spatially smooth even though the latent factors are mutually independent. 
Furthermore, the prediction surface is fairly similar to the true one.

\section{Exact minimaxity of predictive distributions\label{sec:minimax}}
We here provide the theoretical property of the resulting Bayesian predictive distribution based on the synthesis function (\ref{eq:SVCM_Apprx}), by showing that BSPS
is exact minimax under Kullback-Leibler (KL) loss (i.e. desirable under a statistical decision theoretic framework). 

First, we formulate the decision theoretic problem.
Let $\boldsymbol{s}=\left\{ s_{1},\ldots,s_{n}\right\} $
be $n$ observed locations, and $s_{n+1}$ be an unobserved location.
Further, let $y_i$ be the observed data at location $s_i$ and $y_{n+1}$ be a random variable at unobserved location. 
We assume that $y_i$'s and $y_{n+1}$ are the realization of a Gaussian field, and the mean and variance of $y_i$ are denoted by $m(s_i)$ and $R(s_i)$, respectively.
Define $f_{j,i}=f_{j}(s_{i})$, $\varepsilon_{i}=\varepsilon(s_{i})$, and
$\beta_{ji}=\beta_{j}(s_{i})$.
Then, the best approximation model at location $s_{n+1}$ can be written as
\begin{equation}\label{eq:SVCM_3}
\begin{split}
y_{n+1} & =\mu_{n+1}+\varepsilon_{n+1},\ \ \ \varepsilon_{n+1}\sim N(0,\sigma_{n+1}^{2}) \\
\mu_{n+1}&=\beta_{0,n+1}+\sum_{j=1}^{J}\beta_{j,n+1}f_{j,n+1}.
\end{split}
\end{equation}
Hence, the conditional distribution of $y_{n+1}$ given $\f_{n+1}=(f_{1,n+1}\},\ldots,f_{J,n+1})$, $\bbe_{n+1}=(\beta_{1,n+1},\ldots,\beta_{J,n+1})$ and $\sigma_{n+1}$ is $N(\beta_{0,n+1}+\sum_{j=1}^{J}\beta_{j,n+1}f_{j,n+1}, \sigma_{n+1}^2)$, which can be interpreted as the likelihood of model~(\ref{eq:SVCM_3}).
The task is to predict an unobserved location, $s_{n+1}$, after observing $\boldsymbol{y}=(y_{1},\cdots,y_{n})^{\top}$, using the model (\ref{eq:SVCM_Apprx}) given in the main text.

The KL loss regarding the prediction of $y_{n+1}$ is defined as
\begin{equation*}
\mathsf{KL}\left[\phi(\cdot;\mu_{n+1},\sigma_{n+1}^2)\mid q(\cdot)\right]
=\int_{\mathbb{R}}\log\frac{\phi(y_{n+1};\mu_{n+1},\sigma_{n+1}^2)}{q(y_{n+1})}\phi(y_{n+1};\mu_{n+1},\sigma_{n+1}^2) dy_{n+1},
\end{equation*}
where $\phi(x; a,b)$ denotes the density function of a normal distribution with mean $a$ and variance $b$.
The goal of statistical decision theory is to determine a distribution, $q(y_{n+1})$, that is close to the true Gaussian distribution, $N(\mu_{n+1},\sigma_{n+1}^{2})$, in terms of KL loss.
We will show that the Bayesian predictive distribution the model given later, $q^{\ast}(y_{n+1}\mid \boldsymbol{y})$, provides a minimax solution to this problem, namely
$$
\mathbb{E}_{\mu_{Y}}\left[\mathsf{KL}(\phi(\cdot;\mu_{n+1},\sigma_{n+1}^2)\mid q^{\ast}(\cdot|\boldsymbol{y})\right]
=\min_q\max_{\mu_{n+1},\sigma_{n+1}}\mathbb{E}_{\mu_{Y}}\left[\mathsf{KL}(\phi(\cdot;\mu_{n+1},\sigma_{n+1}^2)\mid q(\cdot|\boldsymbol{y}))\right].
$$
We define the invariant decision problem.
The parameter space is $\theta=\left(\mu_{n+1},\sigma_{n+1}\right)$, and the decision space is a space of probability distributions.
Denote the decision function as the predictive distribution, $q\left(y_{n+1}\right)$.
Let the orthogonal group, $\mathcal{O}_{n+1}$, be the group of $\left(n+1\right)\times\left(n+1\right)$ orthogonal
matrices, and describes the positive interval, $\left(0,\infty\right]$, on $\mathbb{R}_{+}$.
Define the group, $G$, as $G=\mathbb{R}_{+}\times\mathcal{O}_{n+1}\times\mathbb{R}$, $c\in\mathbb{R}_{+}$, $A\in\mathcal{O}_{n+1}$ and $F\in\mathbb{R}$.
Define the operation, $g\left(\in G\right)$, to the sample space as
\begin{equation}
g\left[\begin{array}{c}
\boldsymbol{y}\\
y_{n+1}
\end{array}\right]=cA\left[\begin{array}{c}
\boldsymbol{y}\\
y_{n+1}
\end{array}\right]+F
\label{eq:Group}
\end{equation}
and define the operation, $\overline{g}(\in G)$, to the parameter space, as the $(n+1)$th component of the following vectors
\begin{align*}
\overline{g}\left[\begin{array}{c}
\boldsymbol{\mu}\\
\mu_{n+1}
\end{array}\right] & =cA\left[\begin{array}{c}
\boldsymbol{\mu}\\
\mu_{n+1}
\end{array}\right]+F, \ \ \ \ \ 
\overline{g}\left[\begin{array}{cc}
\boldsymbol{\sigma}^2 & 0\\
0 & \sigma_{n+1}^{2}
\end{array}\right] & =c^{-1}A\left[\begin{array}{cc}
\boldsymbol{\sigma}^2 & 0\\
0 & \sigma_{n+1}^{2}
\end{array}\right]A^{\top}.
\end{align*}
Note that $\boldsymbol{\mu}=\left(\mu_{1},\cdots,\mu_{n}\right)^{\top}$, $\boldsymbol{\sigma}^2={\rm diag}\left(\sigma_1^2,\cdots,\sigma_n^2\right)$.
Then define the transformation, $\tilde{g}$, to the probability density, $q\left(y\right)$, as $\tilde{g}q\left(y\right)=q\left(gy\right)$.
Thus,
\[
\tilde{g}q\left(y_{n+1}\right)=q\left(\left(n+1\right)\textrm{th component of }cA\left[\begin{array}{c}
\boldsymbol{y}\\
y_{n+1}
\end{array}\right]+F\right).
\]
The transformation group, $g$, operates on the sample, $(\boldsymbol{y}^\top, y_{n+1})$, transitively.
Thus, for any point, $\check{\boldsymbol{y}}$, on $\mathbb{R}$, there necessarily exists $g\in G$, such that $g\check{\boldsymbol{y}}=\boldsymbol{y}$.
Similarly, $\overline{g}$ transitively operates on the parameter space, $\theta\in\left(\mathbb{R},\mathbb{R}_{+}\right)$.
Such transformation groups, $\left(g,\overline{g},\tilde{g}\right)$, make the statistical decision problem invariant.
Thus, regarding the probability distribution, $N(\mu_{t+1},\sigma_{n+1}^2)$, that follows the best approximation model for the sample, $y_{n+1}$, $N(\overline{g}\mu_{n+1},\overline{g}\sigma_{n+1}^{2})=\tilde{g}N(\mu_{t+1},\sigma_{n+1}^{2})$ holds, and regarding the KL loss, we have
\begin{align*}
\mathsf{KL}\left[\phi(\cdot;\overline{g}\mu_{n+1},\overline{g}\sigma_{n+1}^{2})\mid\tilde{g}q(\cdot|\boldsymbol{y})\right]
=&\int\log\frac{\phi(y_{n+1};\overline{g}\mu_{n+1},\overline{g}\sigma_{n+1}^{2})}{\tilde{g}q\left(y_{n+1}\left|\boldsymbol{y}\right.\right)}\phi(y_{n+1};\overline{g}\mu_{n+1},\overline{g}\sigma_{n+1}^2)dy_{n+1}\\
= & \int\log\frac{\tilde{g}\phi(y_{n+1};\mu_{t+1},\sigma_{n+1}^2)}{q\left(gy_{n+1}\left|\boldsymbol{y}\right.\right)}\tilde{g}\phi(y_{n+1};\mu_{t+1},\sigma_{n+1}^2)dy_{n+1}\\
= &  \ \mathsf{KL}\left[\phi(\cdot; \mu_{t+1},\sigma_{n+1}^2)\mid q(\cdot|\boldsymbol{y})\right],
\end{align*}
which satisfies loss invariance.
From the above, we have defined the $G$-invariant statistical decision problem.

The (non-stochastic) decision space is a set of distributions of $y_{n+1}$, $q(y_{n+1}\mid\theta)$, and the randomized decision function is treated as equivalent to the predictive distribution, $q(y_{n+1}\mid\boldsymbol{y})$.
This is because the randomized decision function is equivalent to integrating the posterior over some part of the decision space $\int_{q(y_{n+1}\mid\theta)\in D}q(y_{n+1}\mid\theta)q(\theta\mid\boldsymbol{y})d\theta$, and the average $\mathsf{KL}$ loss of the randomized decision function, $q(\theta\mid\boldsymbol{y})$, is 
\[
\int\mathsf{KL}\Big[\phi(\cdot;\mu_{n+1},\sigma_{n+1}^{2})\mid q(\cdot|\theta)\Big]
q(\theta|\boldsymbol{y})d\theta
=\mathsf{KL}\Big[\phi(\cdot; \mu_{n+1},\sigma_{n+1}^{2})\mid q(\cdot|\boldsymbol{y})\Big].
\]
As seen from this formulation, the choice of randomized decision function is equivalent to choosing the predictive distribution, $q\left(\left.y_{n+1}\right|\boldsymbol{y}\right)$.
For a predictive distribution to be $G$-invariant, it is required that $\tilde{g}q(y_{n+1}|g\boldsymbol{y})=q(y_{n+1}|\boldsymbol{y})$.

\begin{thm}[\cite{Kiefer_57}]
When the group, $G$, is defined as (\ref{eq:Group}), the minimax solution to the statistical decision problem exists within the solution to the invariant statistical decision problem:

$$
\min_{q}\max_{\theta}\mathbb{E}\left[\mathsf{KL}\big(\phi(\cdot;\mu_{t+1},\sigma_{n+1}^{2})| q(\cdot|\boldsymbol{y})\big)\right]
=\min_{q:\text{G-invariant}}\max_{\theta}\mathbb{E}\left[\mathsf{KL}\big(\phi(\cdot;\mu_{t+1},\sigma_{n+1}^{2})| q(\cdot|\boldsymbol{y})\big)\right].
$$
\end{thm}

\medskip
From this theorem, we can know that the minimax solution can be found within $G$-invariant predictive distributions.
We now construct a predictive distribution that satisfies this condition.
For the spatially-varying coefficient model~(\ref{eq:SVCM_Apprx}), we approximate the parameter function, $\left(\beta_{0}\left(s\right),\cdots,\beta_{J}\left(s\right)\right)$, with Gaussian processes, $\boldsymbol{\beta}_{j}\left(s\right)\sim\textrm{GP}\left(\tau_{j},g_j\right)$ for $j=0,1,\ldots,J$, where $\boldsymbol{\beta}_{j}\left(s\right)$ are independent for each $j$.
Then, the approximation using Gaussian processes (\ref{eq:SVCM_Apprx}) for each observation location can be written as
\begin{align}
&y_{i}=\beta_{0i}+\sum_{j=1}^{J}\beta_{ji}f_{ji}+\varepsilon_{i},\ \varepsilon_{i}\sim N\left(0,\sigma_{i}^{2}\right)\label{eq:SVCM}\\
& (\boldsymbol{\beta}_{j}\mid\tau_{j},g_j)\sim N\left(\boldsymbol{0}_{n},\tau_{j}H\left(g_j\right)\right),\ j=1,\cdots,J, \ \ \ \pi\left(\boldsymbol{\beta}_{0}\right)=1_{\left[-a,a\right]}\left(\boldsymbol{\beta}_{0}\right).\nonumber 
\end{align}
Here, $H(g_j)$ is an $n\times n$-matrix whose $(i,i')$-element being $\kappa_{g_j}(s_{i},s_{i'})$ and $\kappa_{g_j}(s_{i},s_{i^{\prime}})$ is a kernel function such as the exponential kernel, $\kappa_{g_j}(s_{i},s_{i^{\prime}})=\exp(-\|s_{i}-s_{i'}\|/g_j)$.
The conditional distribution of $y_i$ is $N(\beta_{0i}+\sum_{j=1}^{J}\beta_{ji}f_{ji}, \sigma_i^2)$ for $i=1,\ldots,n+1$.
Let $\rho(\tau_{j}),\rho(g_j)$ and $\rho(\sigma_i)$ be priors on $\tau_j$, $g_j$ and $\sigma_i$, respectively. 
Then, the conditional distribution of $(\boldsymbol{y}^{\top},y_{n+1})$ given $(\f_1,\ldots,\f_J)$ ($J$ predictions at $n$ observed locations) and $(f_{1,n+1},\ldots,f_{J,n+1})$ ($J$ predictions at a new location) is expressed as 
\begin{align*}
&g^{\ast}\big(\boldsymbol{y}, y_{n+1} \mid \f_1,\ldots,\f_J,f_{1,n+1},\ldots,f_{J,n+1} \big)
\\
&= \int \prod_{i=1} \phi\Big(y_i; \beta_{0i}+\sum_{j=1}^{J}\beta_{ji}f_{ji}, \sigma_i^2 \Big)\rho(\sigma_i) d\sigma_i
\prod_{j=0}^{J}\pi(\bbe_j, \beta_{j,n+1}|\tau_j, g_j)d\bbe_j d\beta_{j,n+1}\rho(\tau_{j})\rho(g_j)d\tau_{j}dg_j 
\end{align*}

Then conditional (Bayesian predictive) distribution of $y_{n+1}$ given observed $\boldsymbol{y}$ is 
\begin{equation}\label{eqn:Bayes-pred}
\begin{split}
q^{\ast}\big(y_{n+1} \mid \boldsymbol{y}, &\f_1,\ldots,\f_J,f_{1,n+1},\ldots,f_{J,n+1}\big)\\
&=\frac{\int g^{\ast}\big(\boldsymbol{y},y_{n+1} \mid \f_1,\ldots,\f_J,f_{1,n+1},\ldots,f_{J,n+1} \big) d\boldsymbol{y}}
{\int g^{\ast}\big(\boldsymbol{y},y_{n+1} \mid \f_1,\ldots,\f_J,f_{1,n+1},\ldots,f_{J,n+1} \big) d\boldsymbol{y}},
\end{split}
\end{equation}
which will be simply denoted by $q^{\ast}\big(y_{n+1} \mid \boldsymbol{y})$.
The property of $q^{\ast}\big(y_{n+1} \mid \boldsymbol{y})$ can be shown as follows:

\begin{lem}
Given priors $\rho(\tau_{j})=\tau_{j}^{-1} \ (j=0,1,\ldots,J)$  and $ \rho(\sigma_i)=\sigma_i^{-1} \ (i=1,\ldots,n)$, and any choice of prior for $g_j$, the predictive distribution $q^{\ast}\big(y_{n+1} \mid \boldsymbol{y})$ given in (\ref{eqn:Bayes-pred}) is $G$-invariant. 
\end{lem}

\medskip
For the statistical decision problem we are considering, the Bayes solution is the predictive distribution, $q(y_{n+1}|\boldsymbol{y})$, that minimizes the Bayes risk
$$
\mathbb{E}_{\pi\left(\mu,\sigma\right)}\left[\mathbb{E}_{\mu_{Y}}\left[\mathsf{KL}\left(\left.N\left(\mu_{n+1},\sigma_{n+1}^{2}\right)\right|q\left(\left.y_{n+1}\right|\boldsymbol{y}\right)\right)\right]\right],
$$
which is the expectation of the KL risk under the unknown parameters, $\theta=(\mu_{n+1},\sigma_{n+1})$, with priors, $\pi(\mu_{n+1},\sigma_{n+1})$.
We can consider the prior on $\mu_{n+1}$ as the following.
Consider $\pi(\mu_{n+1})$ as a mixture probability distribution of the random variable, $(\beta_{0,n+1},\ldots,\beta_{J,n+1})$, with weight $(f_{1,n+1},\ldots,f_{J,n+1})$.
Then, when $(\beta_{0,n+1},\ldots,\beta_{J,n+1})$ is given as (\ref{eq:SVCM}) and $\rho\left(\tau_{j}\right)=\tau_{j}^{-1}$, $\pi(\mu_{n+1})$ is $\overline{g}$-invariant.
This is because of the following rotational invariance of the Gaussian distribution:
\begin{align*}
\overline{g}\boldsymbol{\mu} 
&=
\left(cA\boldsymbol{\beta}_{0}+F\right)+\sum_{j=1}^{J}cA\boldsymbol{\beta}_{j}\circ\boldsymbol{f}_{j}
=
\boldsymbol{\beta}_{0}+\sum_{j=1}^{J}\boldsymbol{\beta}_{j}\circ\boldsymbol{f}_{j}=\boldsymbol{\mu},
\end{align*}
where $\circ$ denotes the Hadamard product, and we assume $\boldsymbol{\beta}_{0}$ follows a uniform distribution.
Since $(\beta_{0,n+1},\ldots,\beta_{J,n+1})$ is not the target of estimation, $(\beta_{0,n+1},\ldots,\beta_{J,n+1})$ need not be all $\overline{g}$-invariant.
We then obtain the following result:
\begin{thm}\label{prp:minimax}
Construct a parameter function, $(\beta_{0}(s),\cdots,\beta_{J}(s))$, with a Gaussian process, $\boldsymbol{\beta}_{j}(s)\sim\textrm{GP}(\tau_{j},g_j)$ for $j=0,1,\cdots,J$, and assign priors, $\rho(\tau_{j})=\tau_{j}^{-1}$, $\rho(\sigma_i)=\sigma_{i}^{-1}$, and any prior for $g_j$.
Then, the $G$-invariant predictive distribution $q^{\ast}\big(y_{n+1} \mid \boldsymbol{y})$ given in (\ref{eqn:Bayes-pred}) is a Bayes solution and exact minimax in terms of the KL risk. 
\end{thm}

\begin{proof}

To show exact minimaxity, we first define the transformation group that makes the statistical decision problem invariant under the KL risk.
Here, the orthogonal group, $\mathcal{O}_{n+1}$, is the group of $\left(n+1\right)\times\left(n+1\right)$ orthogonal matrices, with $\mathbb{R}_{+}$ representing the positive region, $\left(0,\infty\right]$.
The group, $G$, is
\begin{align*}{1}
G & =\mathbb{R}_{+}\times\mathcal{O}_{n+1}\times\mathbb{R}, \ \ \  
c\in\mathbb{R}_{+},\ \ \ 
A\in\mathcal{O}_{n+1},\ \ \ 
F\in\mathbb{R},
\end{align*}
where the operation $g$ to the sample space of $(\y, y_{n+1})$ is defined as
$$
g\left[\begin{array}{c}
\y\\
y_{n+1}
\end{array}\right]=cA\left[\begin{array}{c}
\y\\
y_{n+1}
\end{array}\right]+F,
$$
and the operation $\overline{g}$ to the parameter space, $\theta=\left(\boldsymbol{\mu},\mu_{n+1},C_{w},\sigma^{2}\right)$, is defined as
\begin{align*}
\overline{g}
\left[\begin{array}{c}
\boldsymbol{\mu}\\
\mu_{n+1}
\end{array}\right]
& =cA
\left[\begin{array}{c}
\boldsymbol{\mu}\\
\mu_{n+1}
\end{array}\right]
+F, \ \ \ \ 
\overline{g}C=cACA^{\top},\ \ \ \ 
\overline{g}\sigma^{2}=c\sigma^{2}A^{\top}A,
\end{align*}
where $C$ is a $(n+1)\times (n+1)$ covariance matrix. 
The transformation, $\tilde{g}$, to the probability density, $q\left(y\right)$, is defined as $\tilde{g}q\left(y\right)=q\left(gy\right)$.
The transformation group, $\left(g,\overline{g},\tilde{g}\right)$, operates transitively on the sample, $(\y, y_{n+1})$, and the parameter space.

The statistical decision problem is invariant under the transformation group, $\left(g,\overline{g},\tilde{g}\right)$.
Thus, for the sample, $y_{n+1}$, that follows a probability distribution, $p_{\theta}^{*}$, it holds that $p_{\overline{g}\theta}^{*}=\tilde{g}p_{\theta}^{*}$, which entails that the loss is invariant under the KL loss,
\begin{align*}
\mathsf{KL}\left(p_{\overline{g}\theta}^{*}\left|\tilde{g}q\right.\right) & =\int\log\frac{p_{\overline{g}\theta}^{*}\left(y_{n+1}\right)}{\tilde{g}q\left(y_{n+1}\left|\boldsymbol{y},\boldsymbol{f}_{n+1}\right.\right)}p_{\overline{g}\theta}^{*}\left(y_{n+1}\right)dy_{n+1}\\
= & \int\log\frac{p_{\theta}^{*}\left(g(s_{n+1})\right)}{q\left(g(s_{n+1})\left|\boldsymbol{y},\boldsymbol{f}_{n+1}\right.\right)}p_{\theta}^{*}\left(g(s_{n+1})\right)dy_{n+1}=\mathsf{KL}\left(p_{\theta}^{*}\left|q\right.\right).
\end{align*}
The group, $\left(g,\overline{g},\tilde{g}\right)$, is an amenable group, which satisfies the Hunt-Stein condition (Bondar and Milnes, 1981).
The conditions for minimaxity in Kiefer (1957) is thus satisfied.
Therefore, the minimax solution for the given statistical decision problem exists in the solution of the invariant statistical decision problem:
\[
\min_{q}\max_{\theta}{\rm E}_y\left[\mathsf{KL}\left(p_{\theta}^{*}\left|q\right.\right)\right]
=\min_{q:g\textrm{-invariant}}\max_{\theta}{\rm E}_y\left[\mathsf{KL}\left(p_{\theta}^{*}\left|q\right.\right)\right],
\]
where the minimum is taken over a class of $g$-invariant distributions.

From this argument, the best invariant predictive distribution from the class of $g$-invariant distributions is the minimax solution out of all probability distributions.
The best invariant predictive distribution is $\overline{g}$-invariant, i.e., the Bayes decision based on the prior, $\rho$, that satisfies $\rho\left(\overline{g}\boldsymbol{\beta}\right)=\rho\left(\boldsymbol{\beta}\right)$,
$\rho\left(\overline{g}\sigma\right)=\rho\left(\sigma\right)$ provides the best invariant solution (Zidek, 1969).
Under KL risk, the Bayesian predictive distribution under a $\overline{g}$-invariant
prior is the best invariant solution (Komaki, 2002).
For the BSPS model with Gaussian processes for $\beta_1(s),\ldots,\beta_J(s)$, if we use the following $\overline{g}$-invariant prior distributions 
\begin{equation}
\rho\left(\bbe_{0}\right)=1_{\left[-b,a\right]^n}\left(\bbe_0\right), \ \ \
\rho\left(\sigma\right)=\frac{1}{\sigma}, \ \ \ 
\rho\left(\tau_{j}\right)=\frac{1}{\tau_{j}}, \ \ \  j=1,\cdots,J,
\label{eq:invariant-prior}
\end{equation}
for some positive constants, $a$ and $b$.
Here, $1_{\left[-b,a\right]^n}\left(\bbe_0\right)$ is an indicator function, where it is 1 when $\boldsymbol{\beta}_{0}$ is in the region, $\left[-b,a\right]^n$, and 0 otherwise.
Therefore, the predictive distribution under the prior (\ref{eq:invariant-prior}), and all predictive distributions that dominate it, is a minimax solution.

\end{proof}

\vspace{0.5cm}

\end{document}